\documentclass[lettersize,journal]{IEEEtran}
\usepackage{amsmath,amsfonts,amssymb}
\usepackage{algorithmic}
\usepackage{algorithm}
\usepackage{array}
\usepackage[caption=false,font=normalsize,labelfont=sf,textfont=sf]{subfig}
\usepackage{textcomp}
\usepackage{stfloats}
\usepackage{url}
\usepackage{verbatim}
\usepackage{graphicx}
\usepackage{cite}
\usepackage{multirow}
\usepackage{xcolor}
\usepackage[most]{tcolorbox}
\usepackage{booktabs}
\usepackage{tabularx}
\usepackage{hyperref}
\begin{document}

\title{AgentServe: Algorithm-System Co-Design for Efficient Agentic AI Serving on a Consumer-Grade GPU}

\author{Yuning Zhang,~\IEEEmembership{Student Member,~IEEE,}~Yan Yan,~Nan Yang,~\IEEEmembership{Member,~IEEE},~and~Dong Yuan,~\IEEEmembership{Member,~IEEE} \thanks{This work is supported by the Australian Research Council (ARC) linkage project (Grant Number: LP220200893).
\par Y. Zhang, Y. Yan and D. Yuan are with the School of Electrical and Computer Engineering, The University of Sydney, Darlington, NSW 2008, Australia. (Email: yuning.zhang1@sydney.edu.au; yyan0244@uni.sydney.edu.au; dong.yuan@sydney.edu.au). \par N. Yang is with Western Sydney University, Sydney, NSW 2116, Australia, and University of Southern Queensland, Toowoomba, QLD 4350, Australia (email: n.yang@westernsydney.edu.au; nan.yang@unisq.edu.au). \par The corresponding author is D. Yuan.}
}
\markboth{Journal of \LaTeX\ Class Files,~Vol.~14, No.~8, August~2021}{Shell \MakeLowercase{\textit{et al.}}: A Sample Article Using IEEEtran.cls for IEEE Journals}

\maketitle

\begin{abstract}
Large language models (LLMs) are increasingly deployed as AI agents that operate in short reasoning–action loops, interleaving model computation with external calls. Unlike traditional chat applications, these agentic workloads require inference serving systems to balance low latency, stable token emission, and throughput under multiple request arrivals from different AI agents. Recent deployments highlight a shift toward running small language models (SLMs) locally on consumer-grade GPUs, driven by privacy, compliance, and cost constraints. When heterogeneous requests overlap on a single GPU, long prefills and short decodes contend for resources, creating head-of-line blocking that destabilizes interactive performance. By analyzing agent workloads, we observe that their execution naturally separates into \textbf{cold prefills}, which process long system prompts, \textbf{resume prefills}, which append tool outputs to cached contexts, and \textbf{short decodes}, which are latency-critical. This mix intensifies contention compared to conventional chatbot serving. We present AgentServe, a single-GPU serving system that ensures stable multi-agent execution under such conditions by isolating prefills from decodes, applying dynamic budgeting to resume prefills, and allocating GPU resources through pre-established CUDA Green Context slots with adaptive control. Evaluation results show that AgentServe significantly improves latency stability while sustaining competitive throughput, achieving up to 2.8$\times$ TTFT improvement and 2.7$\times$ TPOT improvement over state-of-the-art baselines across different settings.
\end{abstract}


\section{Introduction}

Large language models (LLMs) are increasingly deployed not only as conversational agents but also as AI agents capable of planning, invoking external tools, and integrating retrieval-augmented knowledge sources \cite{openai2022chatgpt,arslan2024survey}. Unlike traditional chat applications where a user issues a long prompt and waits for a relatively slow generation, agentic systems \cite{yao2023react} operate in \textbf{short reasoning and action loops} that interleave model computation with frequent external calls, thereby creating new demands on inference serving systems such as low latency, high throughput, and strict output consistency \cite{schick2023toolformer, belcak2025small}. Recent practice highlights a shift toward deploying small language models (SLMs) locally on consumer GPUs rather than relying solely on cloud-hosted LLMs, driven by privacy and data residency requirements that keep inputs and logs on-site \cite{belcak2025small, act2024eu}. As a result, the consumer-level GPU multi-agent server has emerged as a default deployment unit in enterprises, robotics control centers, and vehicular systems, balancing efficiency, cost, and compliance with governance requirements \cite{belcak2025small,durante2025interactive,yang2023foundation}. In this work, we focus on a practical deployment regime in which tool-augmented agents run SLMs on a single consumer-grade GPU. Our objective is to improve serving stability and efficiency in this setting, rather than to claim that local SLM deployments can replace large cloud models for all tasks.

To investigate efficient deployment of small language models (SLMs) on a single GPU while serving multiple agents simultaneously, we begin with the basic principle of autoregressive inference. Each session consists of two phases: a prefill, where the model processes the entire input sequence with full attention to produce the first token, and a decode, where tokens are generated incrementally using the cached key-value states. Prefill is compute-intensive and grows with prompt length, while decode is lightweight per token but highly sensitive to delays because the smoothness of interaction depends on regular token emission.

\begin{figure}[t]
  \centering
  \includegraphics[width=\linewidth]{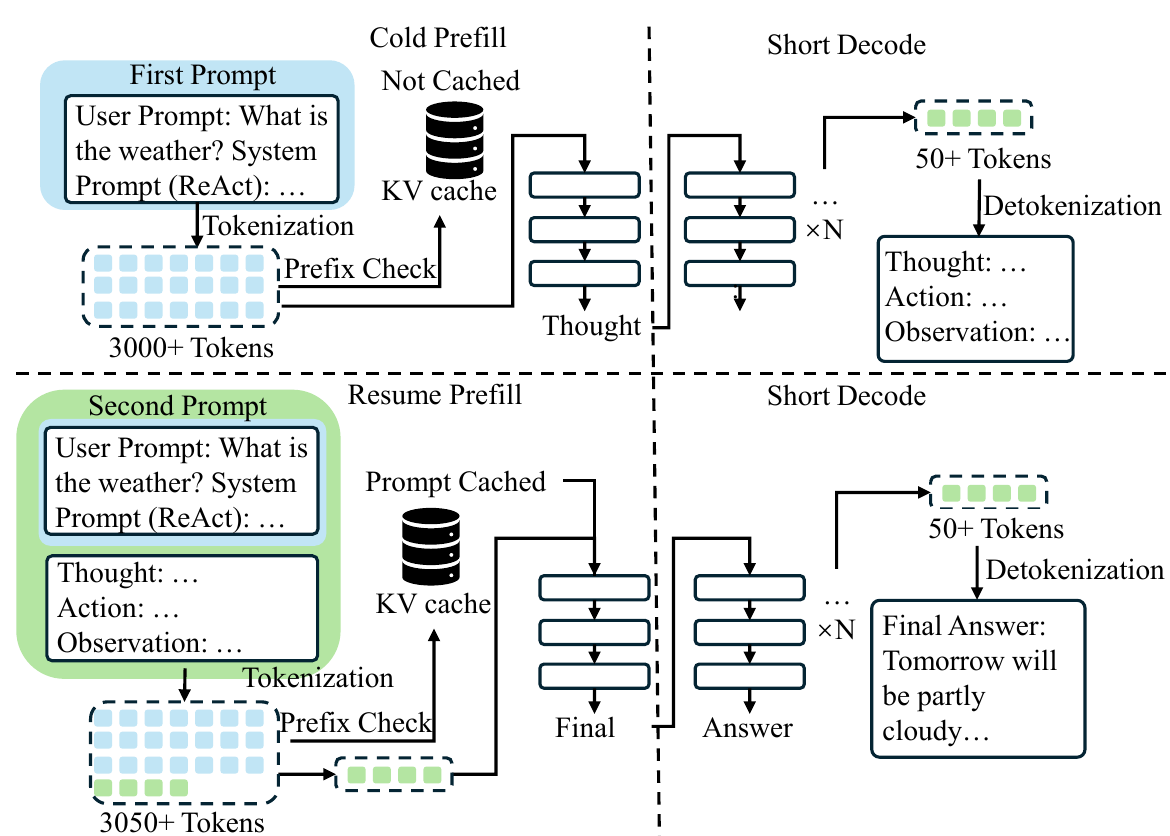}
  \caption{Illustration of cold prefill, resume prefill, and short decode cycles in agent workloads with prefix caching.}
  \label{fig:moti_A}
\end{figure}

Compared with traditional chatbot workloads that involve loosely structured prompts and long decodes, agent workloads are shaped by strict system prompts and structured outputs, which amplify the imbalance between prefills and decodes. We observe that in these workloads, the initial prefill consumes a long system prompt without prefix caching\cite{pan2025marconi}, which monopolizes compute and memory resources. Subsequent prefills append tool outputs or steering text to the cached context, which further adds to interference. In contrast, the decode phase in agent workloads is constrained by the requirement for structured outputs. This constraint leads to decode lengths that are short and relatively stable. Because agents rely on these short decodes to complete each tool invocation before progressing, even minor stalls can break the rhythm of token emission and cascade into longer task latencies. The result is head-of-line (HoL) blocking, where a small number of expensive prefills inflate time-to-first-token (TTFT) for new requests and increase time-per-output-token (TPOT) for ongoing streams. In this paper, we define \textbf{cold prefill} as the long uncached system prompt, \textbf{resume prefill} as cached-context extensions with tool outputs, and \textbf{short decode} as the stable structured generation phase. Figure \ref{fig:moti_A} shows the agent inference workflow and the interaction of these components.

Existing serving systems typically mitigate this issue using either prefill--decode (PD) disaggregation or chunked prefill. Frameworks such as vLLM, SGLang, and DistServe \cite{kwon2023efficient,NEURIPS2024_724be447,298687} were primarily developed for throughput-oriented chatbot-style serving, often with stronger hardware or multi-engine assumptions. In distributed settings, PD disaggregation amortizes KV transfer and orchestration overhead across many GPUs. In single-GPU adaptations, process separation still introduces runtime overhead and does not always guarantee strict decode isolation. Chunked prefill can reduce head-of-line blocking when decodes are long, but in agent loops decodes are often very short, so frequent chunk boundaries still perturb token emission. Therefore, the core challenge in our setting is not simply doing PD disaggregation on one GPU, but controlling interference among \textbf{cold prefills}, \textbf{resume prefills}, and \textbf{short decodes} under tight single-device constraints.

We present \textbf{AgentServe}, a serving system designed for this single-GPU agent regime through an algorithm and system co-design. On the algorithmic side, AgentServe performs phase-aware request classification and TPOT-driven scheduling: cold prefills are isolated, resume prefills are admitted under a dynamic budget, and decodes receive protected resources to preserve stream stability. On the system side, AgentServe uses pre-established CUDA Green Context\cite{nvidia_cuda_green_contexts} slots and lightweight shared-memory coordination to enforce isolation within a single engine, avoiding costly inter-engine KV transfers. Together, these mechanisms maintain stable token emission and competitive utilization in consumer-GPU deployments.

In summary, our paper makes the following contributions:  

\begin{itemize}
\item We propose a resource-aware scheduling algorithm that combines request isolation with TPOT-driven adaptation. It constrains prefill interference and protects latency-critical decodes. We further provide a profile-aware competitive-ratio analysis against an SLO-feasible offline optimum, which bounds prefill-throughput loss under decode SLO constraints.

\item We design a resource management mechanism using pre-established CUDA Green Context slots and lightweight shared-memory coordination, enabling efficient single-engine resource partitioning with low runtime overhead.

\item We present AgentServe, a single-GPU inference serving system for concurrent tool-augmented SLM agents on consumer GPU, co-optimizing scheduling and system-level resource control. Our target setting is local agent workflows with structured tool use and short interaction cycles, rather than arbitrarily compute-intensive general agents. Comprehensive evaluation across multiple models and GPUs shows that AgentServe improves TTFT by up to 2.8$\times$ and TPOT by up to 2.7$\times$ over state-of-the-art baselines, while ensuring stable low-latency serving under multi-agent concurrency.
\end{itemize}
\section{Background and Motivation}
\subsection{Fundamentals of Autoregressive Inference}
Autoregressive language models generate tokens sequentially, and their inference process can be divided into two stages: prefill and decode. In the prefill stage, the model consumes the entire input sequence with full-sequence attention to build the key-value (KV) cache, a quadratic-complexity operation that determines the time-to-first-token (TTFT). Once the first token is produced, the process enters the decode stage, where subsequent tokens are generated incrementally using the cached prefix. This reduces each step to linear complexity and makes decode lighter than prefill, though its performance governs the smoothness of interaction. The latency of this stage is often measured by the time-per-output-token (TPOT), reflecting how quickly responses are streamed. As illustrated in Figure~\ref{fig:moti_A}, an inference session typically starts with a long and expensive cold prefill (processing the system prompt and user query), followed by multiple resume prefills triggered when short fragments such as tool outputs or retrieval results are appended. Each resume prefill is then followed by short decodes, many of which are function calls or routing tokens rather than long passages of text.

To reduce repeated computation, prefix caching \cite{kwon2023efficient} is widely adopted. The KV cache built in the initial cold prefill can be reused across subsequent steps, avoiding recomputation of the entire input. While this optimization reduces redundant computation, it also couples prefills and decodes to the same cache memory, creating potential contention when multiple interactions are processed in parallel.

\textbf{Motivation.} The asymmetry between heavy prefills and lightweight but latency-sensitive decodes becomes problematic when multiple agents share a single GPU. A large prefill kernel may occupy compute and memory resources for an extended period, delaying concurrent decodes that must sustain a stable token emission rate. In agent workflows, this effect is amplified because each session repeatedly alternates between resume prefills and short decodes: a decode stall not only delays current output, but also postpones the next tool invocation and the following inference round. As a result, small per-step interference can accumulate into large end-to-end interaction delays. Therefore, the focus here is a serving-systems question: how scheduling and isolation should be designed under this execution pattern.

\subsection{From Chatbots to Agents}
Traditional chatbot serving typically involves an initial prefill followed by relatively long decodes. This pattern motivates the conventional two-stage abstraction, where prefills are treated as a single class of work. In contrast, agents rely on highly structured system prompts. These prompts encode tool specifications, function signatures, parameter schemas, orchestration rules, and retrieval-augmented passages~\cite{yao2023react,wang-etal-2023-plan}. As shown in Figure~\ref{fig:moti_A}, this makes the cold prefill at the beginning of an agent session very long and computationally intensive. Once the system prompt is cached, the agent alternates between resume prefills, which integrate tool outputs or additional instructions, and very short decodes. Many of these decodes produce only function calls or routing tokens.
In an agent session, the first model entry processes a long uncached prefix containing the system prompt, tool descriptions, orchestration rules, and user query. We call this blocking-prone phase \textbf{cold prefill}. Later model re-entries append tool observations, retrieval snippets, or steering text while reusing cached KV states. We call these frequent cached-context extensions \textbf{resume prefills}. The subsequent generations are often function calls, routing decisions, or short final answers, which we call latency-critical \textbf{short decodes}.

Table~\ref{tab:phase-comparison} summarizes the distinction. These are not new model operations. They are scheduling roles induced by agent execution.

\begin{table}[t]
\centering
\caption{Comparison between conventional inference phases and agent-specific scheduling phases.}
\label{tab:phase-comparison}
\footnotesize
\setlength{\tabcolsep}{2.5pt}
\renewcommand{\arraystretch}{1.12}
\begin{tabularx}{\columnwidth}{
  >{\raggedright\arraybackslash}p{0.16\columnwidth}
  >{\raggedright\arraybackslash}X
  >{\raggedright\arraybackslash}X
  >{\raggedright\arraybackslash}p{0.18\columnwidth}
}
\toprule
Phase 
& In conventional serving 
& In agent serving 
& Required control \\
\midrule
Cold prefill
& Generic prefill before decode
& Long uncached system/tool prompt at session start with high HoL-blocking risk
& Isolate long prefills \\
\midrule
Resume prefill
& Same generic prefill class
& Cached-context extension after tool outputs or retrieval results, short but frequent
& Budgeted admission \\
\midrule
Short decode
& Generic decode, often long enough to amortize scheduling perturbations
& Function call, routing decision, or short structured output where delay blocks the next tool step
& Protect TPOT cadence \\
\bottomrule
\end{tabularx}
\end{table}

\textbf{Motivation.} Unlike chatbots, where decoding dominates computation and throughput optimizations are often sufficient, agents exhibit a fundamentally different workload profile. Prefills dominate computational cost, while decodes dominate latency perception: even small disruptions in decoding break the regular emission of tokens, and because tool invocations are sequential, such delays cascade into much longer end-to-end latencies. This prefill--decode imbalance is further magnified in agents, where cold prefills are long, resume prefills are frequent, and decodes are short but highly sensitive to interference. Consequently, the key design objective shifts from maximizing aggregate token throughput alone to preserving decode regularity under mixed-phase contention.
Thus, treating all prefills uniformly is too coarse: cold prefills should be isolated, while short resume prefills can be admitted when decode pressure is low.
In this paper, we target the \emph{serving-system} problem in local-agent pipelines on consumer GPUs. We do not assume that one SLM can solve every task end-to-end, but rather optimize the deployment regime where tool use and decomposition make local execution practical.

\subsection{Multiple Agents Single-GPU Serving}
\begin{figure}[t]
  \centering
  \includegraphics[width=0.9\linewidth]{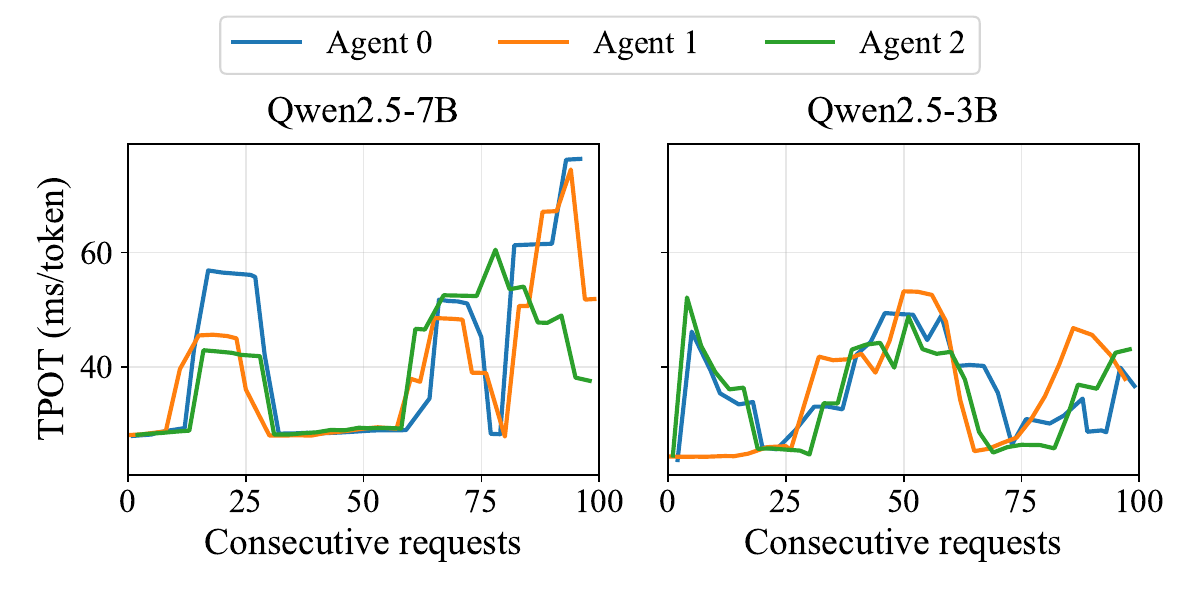}
  \caption{Time per output token (TPOT) of Qwen2.5-7B and Qwen2.5-3B with three concurrent agents on a single RTX A5000 GPU. Cold prefills introduce long kernels that block concurrent decodes, causing visible spikes in token emission latency.}
  \label{fig:TBT}
\end{figure}

When prefills and decodes run together as a mixed workload, head-of-line blocking arises: long cold prefills monopolize compute and memory bandwidth, delaying lightweight but time-critical decodes. This inflates the time-to-first-token (TTFT) for new requests and the time-per-output-token (TPOT) for ongoing ones. In agent workloads, where prefills are longer and decodes shorter, even a few cold prefills can stall many decodes, disrupting the emission heartbeat and compounding interactive delays across agents. Figure~\ref{fig:TBT} shows this temporal effect directly, with sharp TPOT spikes appearing when heavy prefills overlap with active decode streams.

\begin{figure}[t]
  \centering
  \includegraphics[width=0.95\linewidth]{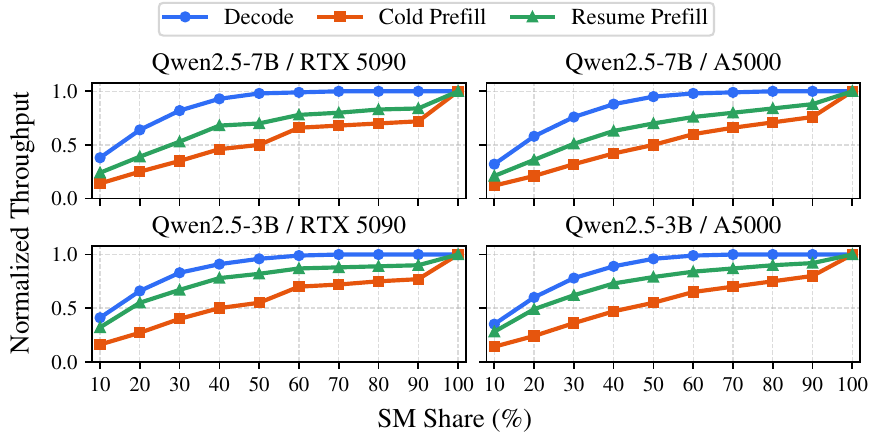}
  \caption{Normalized throughput versus SM share for decode, cold prefill, and resume prefill on Qwen2.5-7B and Qwen2.5-3B. Decode throughput rises quickly at low SM shares and saturates earlier, while prefill throughput increases more gradually.}
  \label{fig:sm-share}
\end{figure}

Complementing this temporal view, Figure~\ref{fig:sm-share} profiles the throughput response to CUDA Streaming Multiprocessors (SMs) allocation for Qwen2.5-7B and Qwen2.5-3B on both RTX A5000 and RTX 5090~\cite{yang2025qwen251mtechnicalreport}. Across both models and both GPUs, decode throughput increases rapidly at low SM shares and saturates earlier than prefill throughput. Cold prefill throughput rises more gradually, while resume prefill remains between decode and cold prefill. This indicates different marginal gains across phases: at low SM shares, decode gains are steep and latency-critical. After the saturation knee, extra decode SMs bring diminishing returns, and allocating residual SMs to prefills is more efficient.

This behavior explains why mixed prefill and decode execution is unstable in multi-agent serving and why fine-grained phase separation is necessary. Long prefills can temporarily dominate SMs and push decode into the steep low-share regime, where token emission quality degrades quickly. AgentServe therefore combines SM-granular decode reservation with budgeted resume-prefill admission, so decode remains near its efficient region while leftover capacity is reclaimed by prefills. Compared with coarse chunking or static partitioning, this fine-grained disaggregation better stabilizes TPOT without sacrificing throughput. Figure~\ref{fig:sm-share} also provides the empirical basis for the profile-aware assumptions used later in the competitive-ratio analysis. 

These observations explain why existing mechanisms are insufficient: a system that treats all prefills uniformly cannot isolate expensive cold prefills, admit lightweight resume prefills for utilization, and protect short decodes at the same time.

PD disaggregation~\cite{298687} separates prefill from decode, but the prefill side still mixes cold and resume prefills. On a single consumer-grade GPU, process-separated variants also incur coordination and memory-bandwidth overhead~\cite{cuda}, and separation alone does not decide which prefills are safe to co-schedule.

Chunked prefill~\cite{298679} slices long prefills and interleaves chunks with decode, but agent decodes are often too short to amortize frequent chunk boundaries. A delayed function call or routing token can postpone the next tool step.

Generic GPU sharing mechanisms such as CUDA MPS and CUDA multi-stream execution~\cite{mps,cudastream} provide concurrency but not semantic phase protection. Recent single-GPU PD studies~\cite{shi2025nexus,hong2025semi,cui2025optimizing} reduce communication overhead, but do not explicitly model the cold-prefill/resume-prefill distinction in tool-augmented agent loops.

AgentServe instead isolates cold prefills, admits resume prefills under a dynamic token budget, and protects short decodes on consumer GPUs.

\textbf{Motivation.}
These observations motivate lightweight, phase-aware scheduling on a single consumer GPU. Rapid phase transitions and bursty decode arrivals limit the effectiveness of static partitioning, while multi-engine designs introduce coordination overhead. An effective solution must distinguish cold and resume prefills, protect latency-sensitive decodes, and adapt resource allocation to changing workload demands.
\section{AgentServe}
\begin{figure*}[h]
  \centering
  \includegraphics[width=0.9\linewidth]{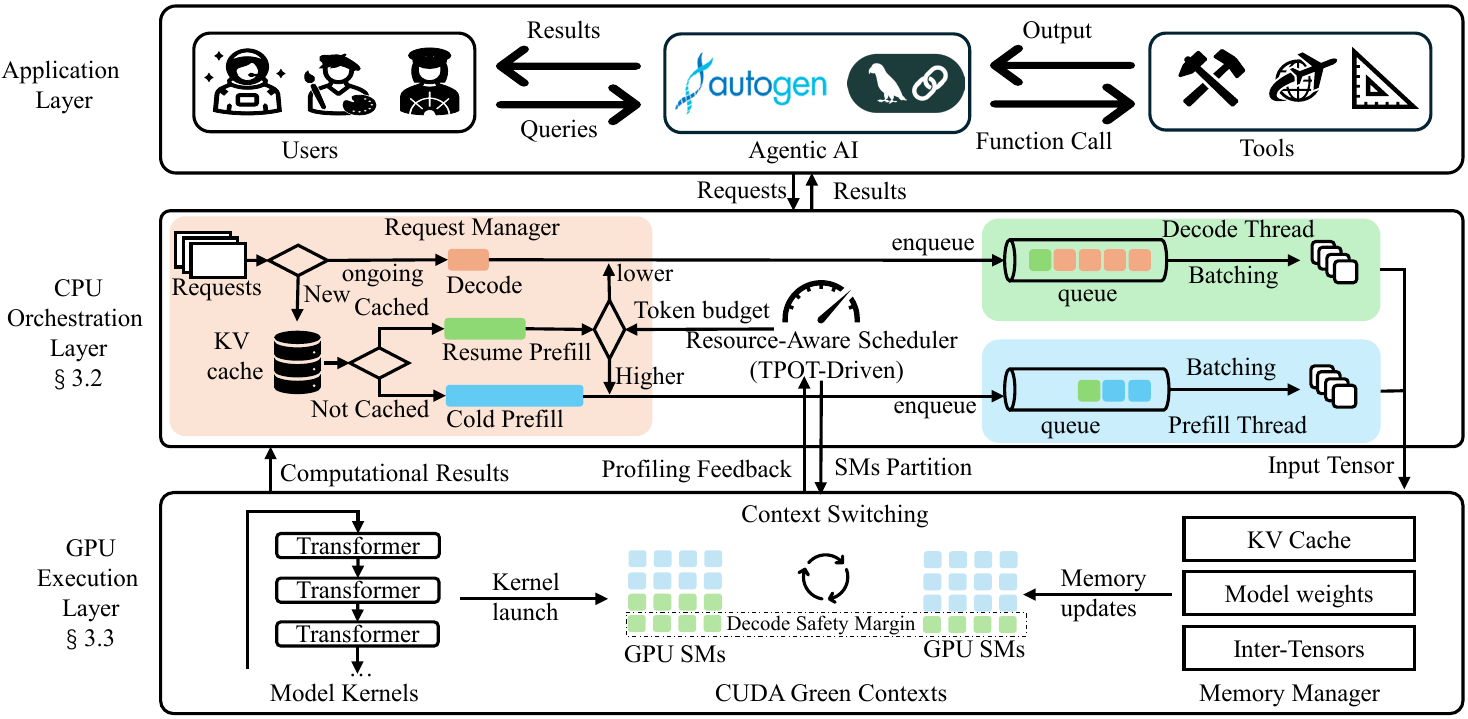}
  \caption{System architecture of AgentServe. The Application Layer connects users and tools, the Orchestration Layer manages requests and resource allocation, and the Execution Layer enforces prefill–decode disaggregation with CUDA Green Contexts and memory management to ensure decode responsiveness in multiple agent settings.}
  \label{fig:method}
\end{figure*}
\subsection{System Architecture Overview}
The design of AgentServe is driven by two key questions that emerged from the background analysis. 
\begin{tcolorbox}[colback=gray!5,colframe=black!50,title=Design Objectives]
\textbf{Algorithmic challenge}: How can cold prefills, resume prefills, and decodes be disaggregated and allocated appropriate resources to avoid head-of-line blocking?\\[4pt]
\textbf{System-level challenge}: How can strict resource isolation be achieved on a single GPU using a single-engine design, thereby preventing contention overhead without resorting to dual-engine disaggregation?
\end{tcolorbox}

To answer these questions, AgentServe introduces a single-GPU architecture that integrates request management, TPOT-driven scheduling, and dynamic resource control. As shown in Figure~\ref{fig:method}, the system is organized into three tightly connected layers.  

\textbf{Application Layer.} Requests enter through agentic AI frameworks such as LangChain, AutoGen, or ToolBench\cite{langchain2023,wu2024autogen,Toolbench}, which mediate between end users and external tools. User queries and tool outputs are formatted into model requests and then forwarded to the serving backend, where the orchestration of prefills and decodes begins.  

\textbf{Orchestration Layer.} This CPU-side layer coordinates request management and scheduling. The Request Manager determines whether an incoming request corresponds to a cold prefill, a resume prefill, or a decode. Long cold prefills, as the primary source of head-of-line blocking, are directed to a dedicated thread and queue, while short prefills may share the decode queue under the token budget. Resume prefills are typically short and are merged with decodes to improve parallelism, unless they exceed a predefined token budget, in which case they are rerouted to the cold prefill queue. Each request is also associated with its KV cache status, enabling reuse of cached states when available. The Resource-Aware Scheduler (TPOT-Driven) dynamically adjusts both the prefill token budget and CUDA streaming multiprocessor (SM) reservations. It enforces a decode-priority policy by reserving a safety margin of SMs for decodes, while scaling resume prefills based on profiling feedback and runtime signals such as token throughput, queue occupancy, and utilization. Allocation decisions are then dispatched to the GPU alongside batched tokens. 

\textbf{Execution Layer.} The GPU-side layer enforces phase disaggregation and resource isolation. Prefill and decode kernels are launched by dedicated CPU threads and executed within CUDA Green Contexts, which allow fine-grained SM partitioning and context switching. This ensures that decodes maintain uninterrupted priority access to compute and memory bandwidth even under heavy load. The Memory Manager maintains coherence between KV caches, model weights, and intermediate tensors, so that prefills can safely reuse cached states without inconsistency. Together, these mechanisms allow prefills and decodes to share GPU resources efficiently while avoiding destructive interference.  

Through this design, AgentServe addresses the two guiding objectives. It stabilizes latency-sensitive decodes by isolating cold prefills and regulating resume prefills under a dynamic budget, and it achieves strict GPU resource isolation through thread-level separation and pre-established CUDA Green Contexts, enabling flexible reallocation without incurring cache transfer overhead. The next section (\S\ref{3.2}) details the scheduling algorithm that resolves the first challenge, while \S\ref{3.3} describes the execution mechanisms that realize the second.

\subsection{Resource-Aware Scheduling}
\label{3.2}

Single-GPU multi-agent serving requires maintaining stable decode latency
while allowing prefill workloads to make progress. In agent workloads,
cold prefills are computationally expensive while decodes are short but
latency-sensitive. If prefills monopolize GPU resources, decode streams
may suffer from large inter-token delays.

To address this challenge, AgentServe introduces a lightweight feedback
scheduler that dynamically regulates two control variables:

\begin{itemize}
\item the resume prefill token budget $B_{\text{prefill}}(t)$, which determines
the maximum length of resume prefills that are allowed to execute together
with decode requests, and
\item the minimum number of streaming multiprocessors (SMs) reserved for
decoding, denoted $R_{\min}(t)$.
\end{itemize}

Both variables are adjusted according to the observed decoding latency,
measured using the metric \emph{time per output token (TPOT)}.

\textbf{TPOT measurement.}
In streaming inference, user-perceived responsiveness is determined by
TPOT. In our system, decoding proceeds in discrete
\emph{decode steps}, where each active stream produces at most one token.

Over a control interval $\Delta t$, the scheduler records

\begin{itemize}
\item the cumulative decode time $\Delta L_{\text{decode}}$, and
\item the number of completed decode steps $\Delta K_{\text{decode}}$.
\end{itemize}

The step-level TPOT is defined as

\[
\text{TPOT}_{\text{step}} =
\frac{\Delta L_{\text{decode}}}{\Delta K_{\text{decode}}}.
\]

When a single stream is active, $\text{TPOT}_{\text{step}}$ equals the
inter-token latency. Under multiple streams, it represents the delay
experienced by each decoding round.

\begin{algorithm}[t]
\caption{TPOT-Driven Resource Scheduling}
\label{alg:tpot}
\begin{algorithmic}[1]
\STATE \textbf{Input:} Queues $Q_D, Q_P$, parameters
$R_{\min}, B_{\text{prefill}}, \theta_{\text{low}}, \theta_{\text{high}},
\Delta_R, \Delta_B, \Delta t$, device $S$
\WHILE{system running}
    \STATE measure $\Delta L_{\text{decode}}, \Delta K_{\text{decode}}$
    \STATE compute $\text{TPOT}_{\text{step}} =
           \Delta L_{\text{decode}}/\Delta K_{\text{decode}}$
    \IF{$\text{TPOT}_{\text{step}} > \theta_{\text{high}}$}
        \STATE $B_{\text{prefill}} \gets
               \max(B_{\min}, B_{\text{prefill}}-\Delta_B)$
        \STATE $R_{\min} \gets
               \min(S, R_{\min}+\Delta_R)$
    \ELSIF{$\text{TPOT}_{\text{step}} < \theta_{\text{low}}$}
        \STATE $B_{\text{prefill}} \gets
               \min(B_{\max}, B_{\text{prefill}}+\Delta_B)$
        \STATE $R_{\min} \gets
               \max(R_{\text{base}}, R_{\min}-\Delta_R)$
    \ENDIF
    \FOR{each $req$}
        \IF{$req.type=$ decode OR $req.len \le B_{\text{prefill}}$}
            \STATE enqueue($Q_D$, $req$)
        \ELSE
            \STATE enqueue($Q_P$, $req$)
        \ENDIF
    \ENDFOR
    \STATE partition SMs:
           $S_{\text{decode}}=R_{\min},
           S_{\text{prefill}}=S-R_{\min}$
    \STATE launch decode jobs from $Q_D$ and prefill jobs from $Q_P$
    \STATE wait for the next control interval $\Delta t$
\ENDWHILE
\end{algorithmic}
\end{algorithm}

\textbf{Algorithm explanation.}
Algorithm~\ref{alg:tpot} implements a feedback control loop that jointly adjusts decode protection and prefill admission. In each control interval, lines 3--4 measure decode workload and compute step-level TPOT. Based on this signal, lines 5--7 enter a protection mode when $\text{TPOT}_{\text{step}} > \theta_{\text{high}}$ by shrinking the allowed resume-prefill length (line 6) and increasing decode SM reservation (line 7), while lines 8--10 perform the complementary relaxation when $\text{TPOT}_{\text{step}} < \theta_{\text{low}}$ by expanding $B_{\text{prefill}}$ and reducing $R_{\min}$. With the updated control variables, lines 12--18 classify incoming requests: decode jobs and short resume prefills are admitted to $Q_D$. Short cold prefills may also enter $Q_D$ if their lengths fall within the current token budget. Longer prefills are redirected to $Q_P$ to avoid blocking latency-critical streams. Finally, lines 19--20 materialize the scheduling decision by partitioning SMs and launching decode/prefill jobs in their corresponding contexts.

\textbf{Competitive-Ratio Analysis under Decode SLO.} We now quantify how much prefill service AgentServe may lose,
relative to an optimal offline scheduler that is subject to the
same decode SLO constraint.
Our goal is not to maximize unconstrained aggregate throughput,
but to characterize the \emph{prefill-throughput retention}
achieved by AgentServe among all schedulers that preserve decode
responsiveness.
This objective matches the design goal of AgentServe, which
explicitly prioritizes TPOT stability through the joint control of
$R_{\min}(t)$ and $B_{\text{prefill}}(t)$.

Let $S$ denote the total number of SMs on the GPU.
Let $\mu_D(R)$ denote the decode throughput when $R$ SMs are
allocated to decode, and let $\mu_C(R)$ and $\mu_R(R)$ denote the
throughputs of cold prefill and resume prefill, respectively.
Motivated by the profiling trends in Figure~\ref{fig:sm-share}, we model the
effective prefill throughput in control interval $t$ as
\begin{equation}
\mu_P(R,t) = \eta_t \mu_C(R) + (1-\eta_t)\mu_R(R),
\label{eq:mu_p}
\end{equation}
where $\eta_t \in [0,1]$ captures the fraction of cold prefill
work in that interval.
This formulation allows the prefill mix to vary over time rather
than assuming a fixed workload composition.

\noindent\textbf{Definition 1 (SLO-feasible scheduler).}
A scheduler $\pi$ is called \emph{decode-SLO-feasible} if for every
control interval $t$, its decode allocation $R_{\pi}(t)$ satisfies
\begin{equation}
\mu_D(R_{\pi}(t)) \ge r_{\min},
\qquad
r_{\min} = \frac{1000}{\tau_{\max}}.
\label{eq:slo_feasible}
\end{equation}
Let $\Pi_{\mathrm{SLO}}$ denote the set of all such schedulers.

\noindent\textbf{Definition 2 (Optimal offline SLO-feasible scheduler).}
Among all schedulers in $\Pi_{\mathrm{SLO}}$, define
\begin{equation}
\pi^\star
=
\arg\max_{\pi \in \Pi_{\mathrm{SLO}}}
\sum_t \mu_P(S - R_{\pi}(t), t)\Delta t.
\label{eq:pi_star}
\end{equation}
Thus, $\pi^\star$ is the offline scheduler that maximizes total
prefill service while preserving the same decode SLO as AgentServe.

\noindent\textbf{Assumption 1 (Monotone profiled scaling).}
The profiled throughput functions $\mu_D(R)$, $\mu_C(R)$, and
$\mu_R(R)$ are non-decreasing in $R$.
This assumption does not require linearity, convexity, or
differentiability. It only states that allocating more SMs does
not reduce throughput.

\noindent\textbf{Assumption 2 (Discrete allocation and bounded overshoot).}
Because the execution layer allocates GPU resources through
pre-established CUDA Green Contexts, decode reservations are drawn
from a discrete set
\begin{equation}
\mathcal{G} = \{g, 2g, \dots, S\},
\label{eq:discrete_set}
\end{equation}
where $g$ is the minimum SM allocation granularity.
Assume the decode SLO is feasible under full-GPU decode allocation,
i.e.,
\begin{equation}
\mu_D(S) \ge r_{\min}.
\label{eq:slo_feasible_full}
\end{equation}
Let
\begin{equation}
R_g^\star = \min \{ R \in \mathcal{G} : \mu_D(R) \ge r_{\min} \}.
\label{eq:rgstar}
\end{equation}
We assume that the runtime controller may conservatively reserve
slightly more decode SMs than $R_g^\star$, but the excess is bounded:
\begin{equation}
R_A(t) \le R_g^\star + \delta,
\qquad
0 \le \delta \le S - R_g^\star,
\label{eq:bounded_overshoot}
\end{equation}
where $\delta$ captures the combined effect of allocation
granularity, control-step size, and finite control lag.

\noindent\textbf{Assumption 3 (Bounded control overhead).}
Let $\varepsilon_{\mathrm{ctx}}(t)$ denote the \emph{equivalent
relative service loss} in interval $t$ caused by context rebinding
and control synchronization, after normalizing any fixed overhead
within the interval by $\Delta t$.
We assume
\begin{equation}
0 \le \varepsilon_{\mathrm{ctx}}(t) \le \bar{\varepsilon} < 1.
\label{eq:eps_bound}
\end{equation}
Accordingly, if the nominal prefill service in interval $t$ is
$\mu_P(R,t)\Delta t$, then the realized service is lower bounded by
\begin{equation}
\bigl(1-\varepsilon_{\mathrm{ctx}}(t)\bigr)\mu_P(R,t)\Delta t
\ge
(1-\bar{\varepsilon})\mu_P(R,t)\Delta t.
\label{eq:eps_service_bound}
\end{equation}

\noindent\emph{Proof.}
By definition, $R_g^\star$ is the smallest element in
$\mathcal{G}$ such that $\mu_D(R) \ge r_{\min}$.
If there existed a scheduler $\pi \in \Pi_{\mathrm{SLO}}$ and an
interval $t$ with $R_{\pi}(t) < R_g^\star$, then by the minimality
of $R_g^\star$ we would have
\[
\mu_D(R_{\pi}(t)) < r_{\min},
\]
which violates the SLO-feasibility condition in
(\ref{eq:slo_feasible}).
Therefore, every SLO-feasible scheduler must satisfy
$R_{\pi}(t) \ge R_g^\star$.
\hfill $\square$

\noindent\textbf{Lemma 2 (Instantaneous upper bound on offline prefill service).}
For any interval $t$, the optimal offline scheduler satisfies
\begin{equation}
\mu_P(S - R_{\pi^\star}(t), t)
\le
\mu_P(S - R_g^\star, t).
\label{eq:lemma2}
\end{equation}

\noindent\emph{Proof.}
By Lemma~1, any SLO-feasible scheduler, including $\pi^\star$,
must allocate at least $R_g^\star$ SMs to decode.
Hence its prefill partition is at most $S - R_g^\star$.
Since $\mu_P(R,t)$ is non-decreasing in $R$ by Assumption~1,
the largest possible prefill throughput under the decode SLO is
achieved when decode uses exactly $R_g^\star$ SMs.
Therefore,
\[
\mu_P(S - R_{\pi^\star}(t), t)
\le
\mu_P(S - R_g^\star, t).
\]
\hfill $\square$

\noindent\textbf{Theorem 1 (Instantaneous competitive ratio).}
Let $W_A(t)$ and $W_{\pi^\star}(t)$ denote the prefill work
completed by AgentServe and $\pi^\star$, respectively, during
interval $t$.
Under Assumptions~1--3, AgentServe satisfies
\begin{equation}
\rho_t
\triangleq
\frac{W_A(t)}{W_{\pi^\star}(t)}
\ge
(1-\bar{\varepsilon})
\cdot
\frac{\mu_P(S - R_g^\star - \delta, t)}
{\mu_P(S - R_g^\star, t)}.
\label{eq:inst_ratio}
\end{equation}

\noindent\emph{Proof.}
By Assumption~2, AgentServe reserves at most $R_g^\star+\delta$
SMs for decode, so its prefill partition satisfies
\begin{equation}
S - R_A(t) \ge S - R_g^\star - \delta.
\label{eq:proof_part1}
\end{equation}
Since $\mu_P(R,t)$ is non-decreasing in $R$, we obtain
\begin{equation}
\mu_P(S - R_A(t), t)
\ge
\mu_P(S - R_g^\star - \delta, t).
\label{eq:proof_part2}
\end{equation}
Accounting for bounded control overhead,
the completed prefill work of AgentServe in interval $t$ is lower
bounded by
\begin{equation}
W_A(t)
\ge
(1-\bar{\varepsilon})
\,\mu_P(S - R_g^\star - \delta, t)\Delta t.
\label{eq:proof_part3}
\end{equation}
On the other hand, by Lemma~2, the offline optimum satisfies
\begin{equation}
W_{\pi^\star}(t)
\le
\mu_P(S - R_g^\star, t)\Delta t.
\label{eq:proof_part4}
\end{equation}
Dividing (\ref{eq:proof_part3}) by (\ref{eq:proof_part4}) yields
(\ref{eq:inst_ratio}).
\hfill $\square$

\noindent\textbf{Corollary 2 (Linearized loss bound).}
If $\mu_P(R,t)$ is Lipschitz continuous on
$[S - R_g^\star - \delta,\; S - R_g^\star]$, i.e., there exists
$L_P(t) \ge 0$ such that for all
$x,y \in [S - R_g^\star - \delta,\; S - R_g^\star]$,
\begin{equation}
\left|\mu_P(x,t) - \mu_P(y,t)\right|
\le
L_P(t)\,|x-y|,
\label{eq:lipschitz}
\end{equation}
then
\begin{equation}
\mu_P(S - R_g^\star - \delta, t)
\ge
\mu_P(S - R_g^\star, t) - L_P(t)\delta,
\label{eq:lipschitz_step}
\end{equation}
and therefore
\begin{equation}
\rho_t
\ge
(1-\bar{\varepsilon})
\left(
1 -
\frac{L_P(t)\delta}{\mu_P(S - R_g^\star, t)}
\right).
\label{eq:linearized_ratio}
\end{equation}

\noindent\textbf{Remark.}
The bound in (\ref{eq:inst_ratio}) shows that AgentServe loses
prefill service relative to the offline optimum only through three
bounded factors: (i) the discrete SM allocation granularity,
(ii) controller conservatism and lag, captured by $\delta$, and
(iii) bounded context-switch/control overhead,
captured by $\bar{\varepsilon}$.
Therefore, as long as the prefill throughput curve around the
operating point is not excessively steep and the runtime overshoot
remains small, AgentServe retains a constant fraction of the
maximum prefill service achievable by any scheduler that also
satisfies the decode SLO.
We empirically examine this stability under larger-model, longer-prefill, and higher-concurrency stress settings in Table~\ref{tab:stress-test}.

\textbf{Stability of the control loop.} The control loop is designed to reject measurement noise rather than react to it: $\mathrm{TPOT}_{\text{step}}$ is aggregated over each interval $\Delta t$ so per-token jitter is averaged before any decision; the thresholds $\theta_{\text{low}}$ and $\theta_{\text{high}}$ form a deadband within which $B_{\text{prefill}}(t)$ and $R_{\text{min}}(t)$ are held unchanged, preventing chatter; and each update is capped by the fixed steps $\Delta B$ and $\Delta R$. Together, interval averaging, hysteresis, and bounded steps keep small fluctuations from inducing oscillatory SM reallocations, as confirmed by Table~\ref{tab:system-overhead} and Table~\ref{tab:stress-test}.
\subsection{Execution System Design}\label{3.3}

The execution layer enforces the SM allocation strategy determined by the Resource-Aware Scheduler and guarantees isolation between prefills and decodes at runtime. The design is centered on two dedicated CPU threads, the Prefill Thread and the Decode Thread, which asynchronously submit kernels to the GPU. This dual-threaded submission ensures that the request pipeline remains non-blocking on the CPU side, enabling asynchronous PD execution within a single engine.

\textbf{Green Context allocation and rebinding.}  
The GPU-side isolation is realized through CUDA Green Contexts\cite{nvidia_cuda_green_contexts}, a feature introduced in recent CUDA versions that enables partitioning of GPU resources at the level of streaming multiprocessors (SMs). Unlike conventional CUDA contexts, which share the full GPU in a time-sliced manner, a Green Context allows the runtime to reserve a fixed subset of SMs exclusively for kernels launched within that context. This provides spatial isolation across workloads, ensuring that latency-sensitive kernels are not preempted or starved by large background jobs. In the Execution Layer, this mechanism enforces a minimum reservation for decoding while leaving the remaining SMs available for prefill.

To enable dynamic partitioning, we pre-establish ten discrete Green Contexts at system initialization, each reserving a fixed fraction of SMs ranging from 10\% to 100\% in increments of 10\%. We choose to create these contexts offline because Green Context construction is substantially more expensive than runtime rebinding. Creating or destroying contexts on demand would introduce control-path overhead directly into latency-sensitive inference. These contexts are created once before inference begins and persist throughout the system’s lifetime, allowing the Execution Layer to switch allocations without incurring the high overhead of repeated context construction. During inference, both the prefill and decode threads periodically report runtime statistics such as TPOT, kernel completion latency, and queue backlog to the Resource-Aware Scheduler. Based on these signals, the Resource-Aware Scheduler computes a target allocation $R_{\min}(t)$ for decoding, and the Execution Layer rebinds the decode thread to the nearest Green Context that guarantees at least $R_{\min}(t)$ SMs, while assigning the complementary context to prefill. For example, if the target allocation is 37\% of SMs, the Execution Layer selects the 40\% context to satisfy the requirement. Because rebinding only involves switching between pre-created contexts, context switching overhead remains below 50 microseconds per rebinding, which is less than 0.1\% of typical decode batch latency, enabling near-continuous control over partitioning while keeping switching overhead negligible.

\textbf{Memory management.}  
Both threads share the same GPU memory pool to avoid KV cache transfers between processes. A Memory Manager coordinates buffer allocation and ensures mutual exclusion. When a prefill completes, its KV cache region is marked read-only and is immediately available to the decode thread without duplication. To prevent race conditions during concurrent access, we employ a combination of CPU-side mutexes and GPU-side cudaEvent synchronization. The mutex guarantees exclusive access when allocating or releasing KV buffers, while cudaEvent enforces ordering between kernel completions in the prefill and decode threads. Together, these mechanisms ensure that decoding never consumes partially written KV states, enabling correct reuse while minimizing synchronization overhead. This design guarantees correctness and prevents race conditions while enabling fine-grained PD disaggregation within a single engine.

\textbf{Thread cooperation.}  
Prefill and decode threads run independently but are tightly synchronized through the Resource-Aware Scheduler’s allocation logic. Prefills are always launched in the Prefill Context with dynamically varying SM shares, while decodes are guaranteed a minimum of $R_{\min}(t)$ SMs in the Decode Context. If profiling reveals that TPOT is rising, the Resource-Aware Scheduler increases the decode floor and the execution layer immediately rebinds the Decode Thread to a larger context. Conversely, when decode demand is light, the Prefill Thread can opportunistically claim more SMs to accelerate prefills. This cooperative mechanism ensures decode responsiveness while maximizing GPU utilization.

\section{Experimental Evaluation}

\subsection{Implementation and Experimental Setup}

\textbf{Implementation.}
We extend llama.cpp(b5943) with CUDA Green Context support, using two dedicated CPU threads for prefills and decodes coordinated by the Resource-Aware Scheduler. At runtime, threads are rebound to pre-created contexts based on TPOT signals, and the KV cache manager ensures safe block reuse across threads without inter-process transfers. Both execution contexts reference the same device-resident KV cache arena, so prefill-to-decode hand-off updates metadata rather than copying KV tensors. Each worker mutates only its own metadata at batch boundaries, and the scheduler validates reuse with per-sequence shadow tokens before falling back to re-prefill when a prefix is unsafe.

\textbf{Hardware Platforms.}  
Experiments are conducted on two servers to study the impact of compute capacity. The first uses an NVIDIA RTX A5000 GPU (64 SMs, 24 GB GDDR6) with an AMD Threadripper PRO 3945WX 12-Core CPU and 128 GB RAM, representing a mid-range edge deployment. The second uses an RTX 5090 GPU (170 SMs, 32 GB GDDR7) with a Threadripper PRO 5955WX 16-Core CPU and 128 GB RAM, representing a next-generation high-performance GPU. Both run Ubuntu 22.04 with CUDA 12.8 and driver 550.54.15. Comparing these platforms allows us to examine how differences in SMs, CPU cores, and memory bandwidth affect system behavior under identical workloads.

\textbf{Models.}  
We evaluate three representative open-source models: Qwen2.5-3B, Qwen2.5-7B\cite{yang2025qwen251mtechnicalreport}, and LLaMA-3-8B\cite{grattafiori2024llama}. These models represent different parameter scales, namely 3B, 7B, and 8B, and they also belong to two distinct architectural families, Qwen and LLaMA. This configuration provides a diverse and realistic testbed for single-GPU agent serving. It enables a thorough assessment of AgentServe across variations in model size and across different architectures, ensuring that the observed latency and stability improvements are not limited to a single model family.

\textbf{Workloads.}  
We construct workloads from ToolBench~\cite{Toolbench}, which provides multi-step, tool-augmented agent tasks involving retrieval, tool invocation, and short reasoning loops. Our goal is to evaluate the serving behavior of \emph{local SLM-based agents on a single consumer GPU}, rather than to emulate arbitrarily compute-intensive general agents such as large coding assistants. In practical on-device deployments, the dominant workload is typically a user-driven agent session with structured tool use, where a long initial prompt is followed by multiple short tool-return updates and brief decoding phases. This pattern matches the cold-prefill / resume-prefill / short-decode structure shown in Figure~\ref{fig:moti_A} and captures the serving bottlenecks that matter most in local agent execution.

Within each agent session, inference proceeds sequentially: a session starts with one cold prefill and then alternates between resume prefills and short decodes as tool outputs are appended. Across sessions, however, we vary the number of concurrent agents from 3 to 6 to emulate realistic multi-agent contention on a single GPU. This setup is intentionally scoped to consumer-hardware local agents: unlike cloud-hosted, highly elastic deployments, such systems are capacity-limited and are not expected to sustain arbitrarily deep reasoning traces or long-form coding workloads. Instead, the key systems challenge is to preserve responsiveness and streaming smoothness under concurrent tool-augmented sessions. 

We evaluate two representative agent paradigms:  
\begin{itemize}
    \item \textbf{ReAct\cite{yao2023react}:} Interleaves reasoning steps with external tool calls. Characterized by frequent resume prefills and extremely short decodes (often function calls or routing tokens). This pattern stresses latency sensitivity, since even small decode delays cascade into longer end-to-end latencies.  
    \item \textbf{Plan-and-Execute\cite{wang-etal-2023-plan}:} Generates an explicit plan before executing actions. Characterized by long cold prefills and medium-length decodes, creating a prefill-intensive workload that stresses the system’s ability to handle large kernels without starving concurrent decodes.  
\end{itemize}  

Together, these two paradigms capture the spectrum of agent workloads: ReAct highlights the risks of decode starvation under frequent short interactions, while Plan-and-Execute emphasizes the computational pressure of cold prefills. In Section \ref{sec:token-dist}, we present token distribution statistics across different workloads and models, demonstrating that agent workloads consistently exhibit the pattern of cold prefill, resume prefill, and short decode. This choice is deliberate: on a single consumer GPU, the most relevant systems question is not whether every agent workload can be supported, but whether common local tool-augmented sessions can be served with stable TTFT and TPOT under contention. Each serving run mixes ReAct and Plan-and-Execute sessions, so concurrent agents vary in execution paradigm, prefill length, resume-prefill frequency, and short-decode behavior.

\textbf{Capacity-aware upper bound.}
We use six concurrent agents as the main A5000 upper bound because this setting is already constrained by both memory capacity and performance saturation. For LLaMA3-8B, the FP16 model buffer occupies 14,315~MiB. With a 4096-token maximum context, KV cache requires about 512~MiB per agent, so six agents add 3,072~MiB and the measured footprint reaches about 23~GB after intermediate tensors, CUDA/runtime workspaces, allocator fragmentation, and framework overhead. Doubling the context to 8192 tokens would raise KV cache to 1,024~MiB per agent and exceed the 24~GB A5000 budget without quantization or reducing context. At this operating point, aggregate throughput also plateaus while TTFT/TPOT tails continue to rise, so adding agents mainly increases queueing and reduces SLO-satisfying goodput rather than improving useful serving capacity. For an eight-agent, 8192-token setting, the model and KV cache alone require 14,315+8$\times$1,024=22,507~MiB, and the observed runtime overhead would raise the footprint to about 28--29~GB. Therefore, we evaluate this stress regime on RTX 5090, whose 32~GB memory provides the required headroom.

\begin{figure*}[t]
    \centering
    \includegraphics[width=0.9\textwidth]{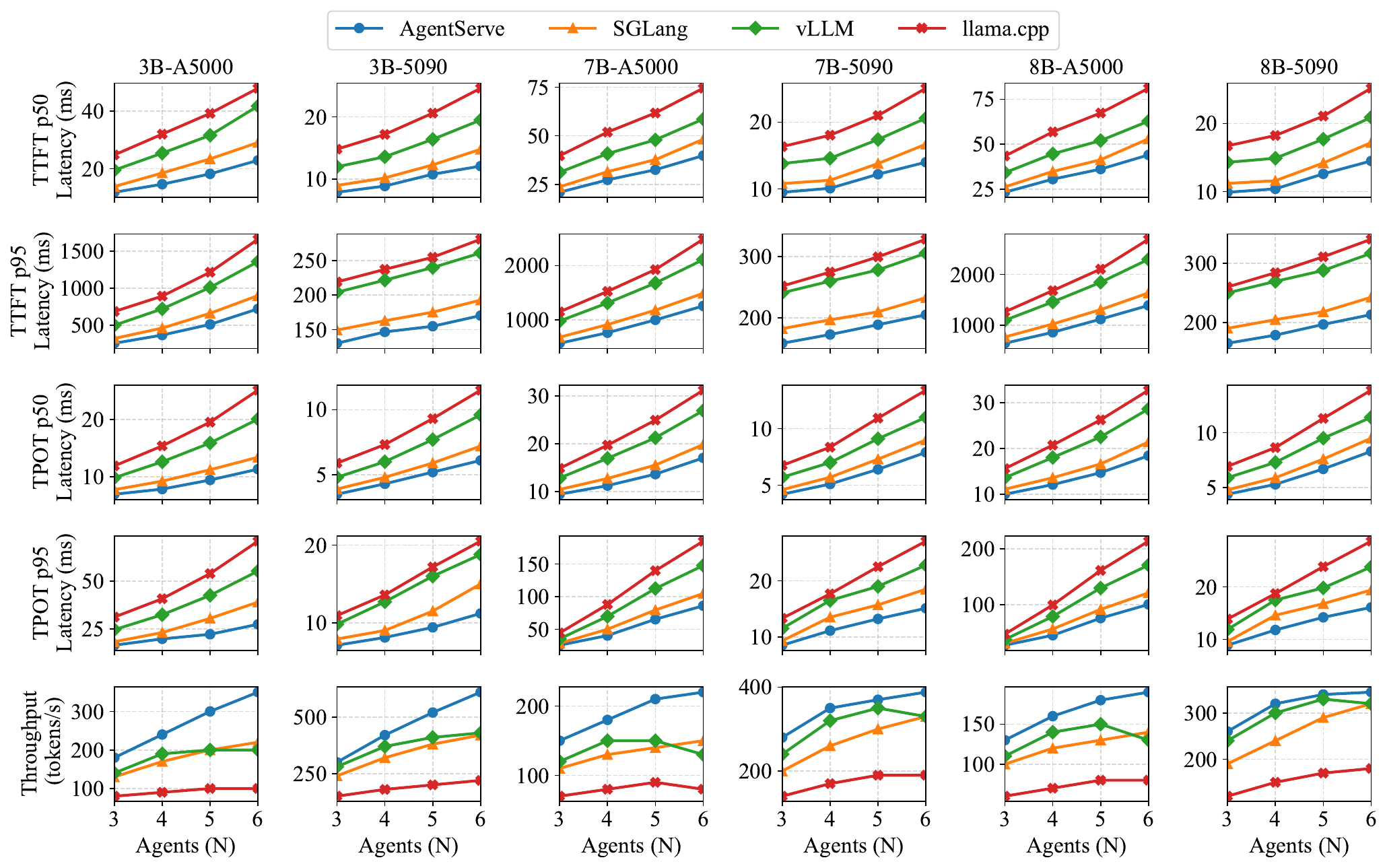}
    \caption{Latency and throughput under different model–device settings. AgentServe consistently achieves the lowest TTFT and TPOT across concurrency levels, while also sustaining competitive throughput compared with baselines. Results are shown for Qwen2.5-3B/7B and Llama-3-8B on both A5000 and 5090 GPUs.}
    \label{fig:latency}
\end{figure*}

\textbf{Baselines.}  
We compare our system against three representative LLM serving engines. These include traditional single-GPU inference engines, chunked prefill, and PD disaggregation. All baselines support prefix caching, covering mainstream single-GPU inference solutions.
\begin{itemize}
\item \textbf{SGLang(0.5.3rc0)}\cite{NEURIPS2024_724be447}: A single-GPU inference engine with explicit prefill--decode disaggregation and shared KV cache, representing PD disaggregation based scheduling solutions.
\item \textbf{vLLM(0.10.2rc1)}\cite{kwon2023efficient}: A throughput-oriented serving engine featuring PagedAttention, continuous batching, and chunk-based prefill, representing mixed prefill--decode optimization.
\item \textbf{llama.cpp(b5943)}\cite{llama_cpp}: A lightweight inference framework optimized for single-device execution, representing a general baseline without specialized scheduling optimizations.
\end{itemize}
Prefix caching is enabled for AgentServe and all baselines whenever supported. We use identical prompts, cached-prefix boundaries, and outputs across systems. When cache-management interfaces vary, we use each system's recommended prefix-cache configuration.

\textbf{Metrics.}  
We focus on four key metrics that capture both latency and efficiency.
First, \textbf{time-to-first-token (TTFT)} measures the delay experienced by new requests before the first output token is produced.
Second, \textbf{time-per-output-token (TPOT)} quantifies the inter-token latency during ongoing decoding. We report the median (p50), which reflects typical user-perceived latency, and the 95th percentile (p95), which highlights tail behavior under load.

Third, \textbf{throughput} measures the aggregate efficiency of the system by counting how many output tokens are generated per unit time across all concurrent sessions. This captures overall serving capacity and complements TPOT, since higher concurrency may increase per-stream token delay while still sustaining high aggregate token generation.

Finally, we have the \textbf{SLO attainment rate}, which reflects the fraction of sessions that meet both latency requirements simultaneously. A session is deemed successful if the TTFT is within its threshold and the TPOT is also within its threshold. This joint definition ensures that the metric captures the complete interactive experience, where users expect both immediate first response and smooth token streaming. The thresholds $\tau_{\text{TTFT}}$ and $\tau_{\text{TPOT}}$ are determined for each model–device pair by profiling their isolated performance and scaling with a constant factor, such that the SLO bounds adapt to different hardware capacities and model sizes while remaining aligned with interactive requirements.
\subsection{Performance Analysis}

\textbf{TTFT.}  
Time-to-first-token directly reflects user-perceived responsiveness, since it determines how quickly the system starts streaming output after receiving a query. As shown in Figure \ref{fig:latency}, AgentServe consistently achieves the lowest TTFT across all settings. Compared to SGLang, median TTFT is typically 1.1--1.3$\times$ faster, while tail latency (p95) shows up to 1.3$\times$ improvement. Against vLLM, the advantage grows to 1.5--1.8$\times$, and relative to llama.cpp the margin is the largest, reaching up to 2.8$\times$ faster in heavy-load conditions. These improvements are preserved on both A5000 and 5090 GPUs, though the absolute latency is naturally lower on the 5090 due to its larger compute and memory capacity. Importantly, under the 7B model with long prompts, prefill contention is more severe, and AgentServe becomes especially effective, further widening the performance gap. This shows that our approach not only lowers average latency but also controls long-tail delays, which are the most disruptive to interactive applications. A similar trend is observed for the Llama-3-8B model. Even with its larger architecture, AgentServe maintains consistent margins over all baselines, showing that our approach generalizes well across model families.

\textbf{TPOT.}  
Time-per-output-token reflects the smoothness of token streaming and the stability of inter-token gaps. As shown in Figure \ref{fig:latency}, AgentServe again demonstrates clear benefits. Median TPOT improves by around 1.1--1.2$\times$ over SGLang, 1.3--1.8$\times$ over vLLM, and more than 1.5$\times$ over llama.cpp. At the p95 tail, AgentServe reduces latency by up to 1.3$\times$ compared with SGLang, nearly 2.0$\times$ compared with vLLM, and up to 2.7$\times$ compared with llama.cpp. The relative improvements are more pronounced on the A5000, where limited compute resources make contention between prefill and decode more visible. On the 5090, stronger hardware lowers overall latency, but AgentServe still maintains consistent margins. These results confirm that resource-aware scheduling effectively stabilizes token emission, ensuring that streaming remains fluid even under varying workloads. With the Llama-3-8B model, the advantage is particularly clear. Even as absolute TPOT increases with model size, AgentServe controls long-tail variability and keeps streaming smooth, underscoring the robustness of resource-aware scheduling across architectures.

\textbf{Throughput.}
Throughput quantifies the overall serving efficiency by measuring the number of tokens generated per second across all concurrent sessions. As illustrated in Figure~\ref{fig:latency}, across all model–device pairs, AgentServe improves throughput by 1.2--1.5$\times$ over vLLM, 1.3--1.5$\times$ over SGLang, and up to 2.0--2.2$\times$ over llama.cpp at high concurrency, while simultaneously preserving latency stability. AgentServe achieves the most stable throughput across different model–device settings, maintaining high aggregate production while effectively bounding token latency. Compared with vLLM, which employs chunked prefill to sustain GPU utilization but experiences higher token-level delay, AgentServe provides comparable or higher throughput while simultaneously preserving latency stability. SGLang benefits from prefill–decode disaggregation, which improves latency, but its process isolation overhead reduces aggregate throughput. Llama.cpp exhibits the weakest scalability, as prefill–decode interference substantially diminishes overall token generation. On larger models such as Llama-3-8B, throughput curves saturate earlier due to heavier prefills. Nevertheless, AgentServe continues to retain a clear margin over baselines, underscoring the robustness of the scheduling strategy across both architectures and hardware capacities.

\subsection{SLO Attainment}
\begin{figure}[t]
    \centering
    \includegraphics[width=0.48\textwidth]{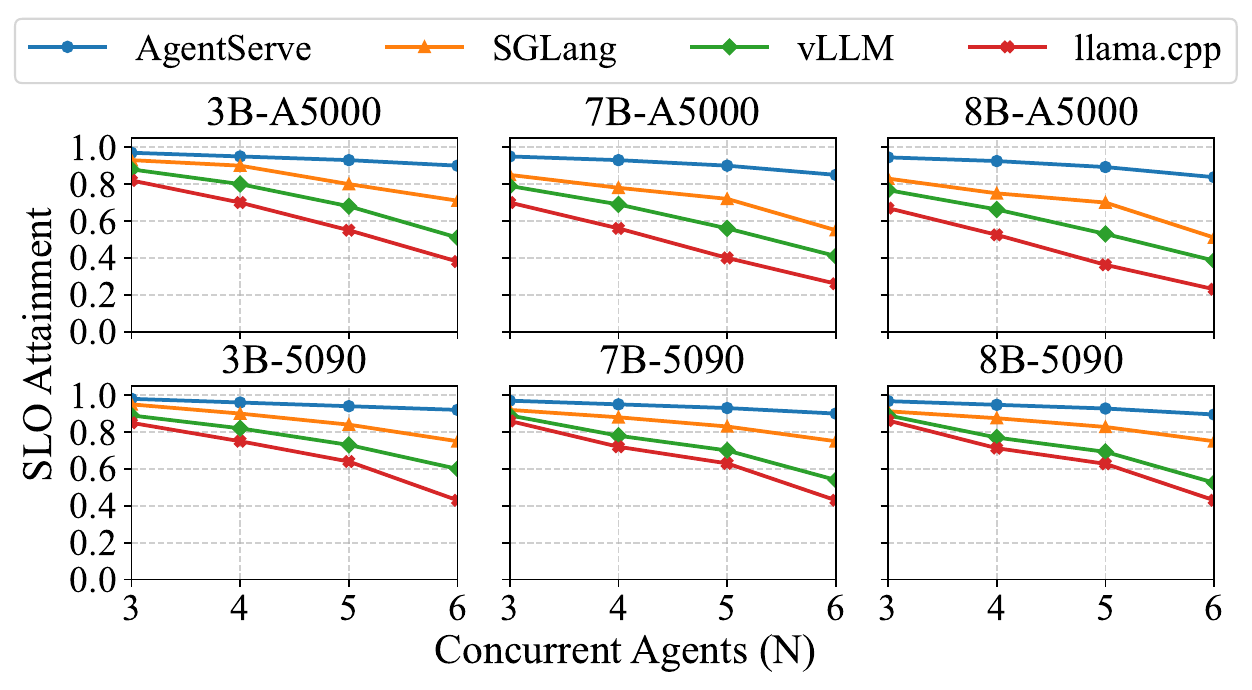}
    \caption{Session-Level SLO Attainment Rate under Varying Agent Concurrency across Models and Devices.}
    \label{fig:slo}
\end{figure}
\textbf{Session-level SLO attainment.} We evaluate the \emph{session-level} SLO attainment rate, defined as the fraction of sessions that satisfy both TTFT and TPOT bounds calibrated for each model--device pair. This joint criterion treats either slow first response or unstable inter-token pacing as a service-level failure.

Figure~\ref{fig:slo} shows SLO attainment across different concurrency levels and model--device settings on both RTX A5000 and RTX 5090. AgentServe achieves the highest compliance, approaching perfect attainment on the RTX 5090 and remaining resilient on the RTX A5000 even for larger models. Its advantage comes from dynamic prefill--decode partitioning, which isolates latency-sensitive decodes and prevents starvation during heavy prefills. In contrast, llama.cpp, lacking prefill--decode optimizations, suffers sharp declines as concurrency grows, with both TTFT and TPOT vulnerable to interference. vLLM mitigates head-of-line blocking with chunked prefill but struggles to balance cold and short resume prefills in agent workloads, leading to suboptimal SLO rates. SGLang, which applies prefill--decode disaggregation but still shares memory, offers better stability than vLLM yet degrades under high concurrency due to contention and lack of strict isolation. On LLaMA3-8B specifically, baseline systems on the A5000 drop quickly once concurrency exceeds four, whereas AgentServe remains stable up to six concurrent agents. On the 5090, AgentServe sustains near-optimal attainment across the tested range. Overall, while existing methods partially address latency issues, only AgentServe's resource-aware disaggregation consistently translates low-level improvements into strong session-level guarantees across hardware and workloads.

\textbf{Workload-specific SLO attainment.}
Since ReAct and Plan-and-Execute have different execution patterns, Table~\ref{tab:workload-slo} breaks down SLO attainment by workload type at six concurrent agents, the highest-concurrency setting in our main evaluation.

\begin{table}[t]
\centering
\caption{Workload-specific SLO attainment at six concurrent agents across all model--device settings.}
\label{tab:workload-slo}
\footnotesize
\setlength{\tabcolsep}{2.2pt}
\renewcommand{\arraystretch}{1.06}
\begin{tabularx}{\columnwidth}{
>{\raggedright\arraybackslash}p{0.19\columnwidth}
>{\raggedright\arraybackslash}p{0.16\columnwidth}
>{\centering\arraybackslash}X
>{\centering\arraybackslash}X
>{\centering\arraybackslash}X
>{\centering\arraybackslash}X
}
\toprule
Setting 
& Workload
& AgentServe 
& SGLang 
& vLLM 
& llama.cpp \\
\midrule
3B-A5000 & ReAct & 0.89 & 0.69 & 0.49 & 0.36 \\
& P\&E & 0.91 & 0.73 & 0.53 & 0.40 \\
\midrule
3B-5090 & ReAct & 0.91 & 0.73 & 0.58 & 0.41 \\
& P\&E & 0.93 & 0.77 & 0.62 & 0.45 \\
\midrule
7B-A5000 & ReAct & 0.84 & 0.53 & 0.39 & 0.24 \\
& P\&E & 0.86 & 0.57 & 0.43 & 0.28 \\
\midrule
7B-5090 & ReAct & 0.89 & 0.73 & 0.52 & 0.41 \\
& P\&E & 0.91 & 0.77 & 0.56 & 0.45 \\
\midrule
8B-A5000 & ReAct & 0.83 & 0.49 & 0.37 & 0.21 \\
& P\&E & 0.85 & 0.53 & 0.40 & 0.25 \\
\midrule
8B-5090 & ReAct & 0.88 & 0.73 & 0.50 & 0.41 \\
& P\&E & 0.91 & 0.77 & 0.55 & 0.45 \\
\bottomrule
\end{tabularx}
\end{table}

AgentServe achieves the highest SLO attainment for both workloads, reaching 0.83--0.93 versus 0.49--0.77 for the strongest baseline. ReAct is slightly harder because frequent tool-call alternation creates more resume prefills and short decodes, but the small gap under AgentServe shows that phase-aware scheduling benefits both patterns. For example, on 7B-A5000, AgentServe improves SLO attainment from 0.53 to 0.84 for ReAct and from 0.57 to 0.86 for Plan-and-Execute compared with SGLang. These results, together with the measured TTFT and TPOT reductions, demonstrate improved serving responsiveness rather than direct reductions in complete agent task latency. Since agent tasks involve repeated model invocations, faster serving can reduce the cumulative inference time across successive steps. However, overall task latency also includes external tool execution and orchestration overhead. When tool execution dominates, serving improvements may translate into smaller relative gains in task completion time. Thus, the actual task-level benefit depends on the proportion of time spent on model inference.

\subsection{Ablation Study and Stress Test}
\begin{figure}[t]
    \centering
    \includegraphics[width=\columnwidth]{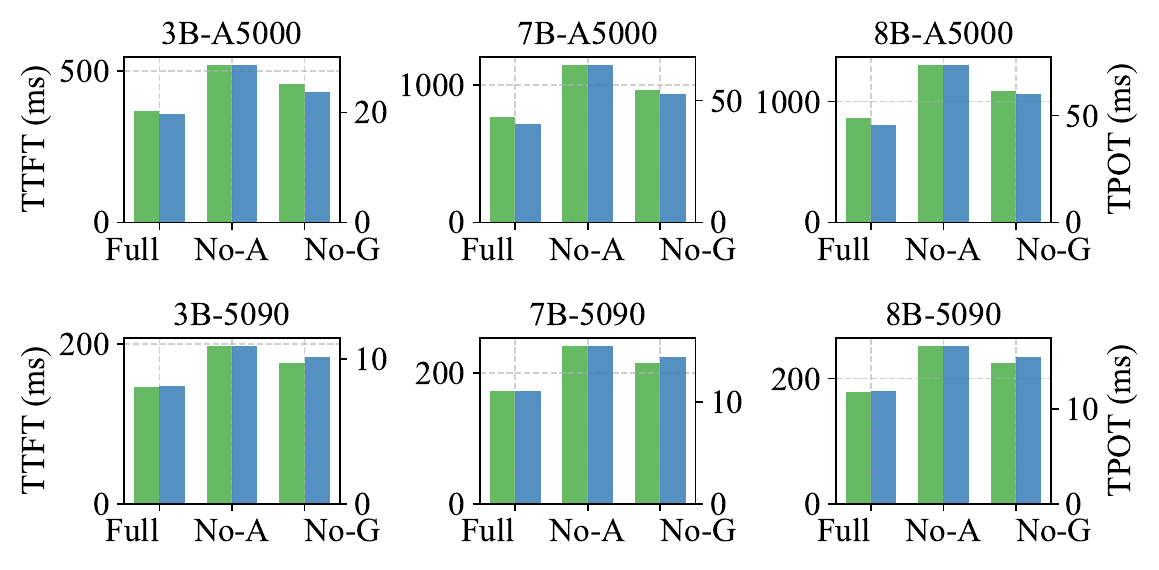}
    \caption{Ablation Study of AgentServe Showing TTFT and TPOT Under Different Scheduling and Resource Isolation Settings.}
    \label{fig:ablation}
\end{figure}
We conduct ablation experiments under four concurrent agents (N=4), reporting p95 tail latency for both TTFT and TPOT. Two simplified variants are compared against the full AgentServe: (\emph{i}) \textbf{No-A}, which disables the resource-aware scheduling and statically partitions SMs between prefills and decodes, removing dynamic adaptation. Both the resume prefill budget $B_{\mathrm{prefill}}$ and decode SM reservation $R_{\min}$ remain fixed during execution. Their values are selected for each model-device configuration based on the SM throughput profiles in Figure~\ref{fig:sm-share} and the workload token distributions in Table~\ref{tab:token-dist}. (\emph{ii}) \textbf{No-G}, which eliminates pre-established CUDA Green Context slots and relies on on-demand stream allocation, providing no decode reservations.

The two adaptive controls address different aspects of interference: the prefill budget regulates the admission of resume prefills alongside decodes, while SM reservation protects the compute capacity available for decoding. Their joint adaptation balances prefill progress and decode responsiveness as workload contention changes. As shown in Figure~\ref{fig:ablation}, both ablations significantly worsen tail behavior. No-A increases TTFT by 15--25\% and degrades TPOT p95 up to 1.4$\times$ due to long prefills starving decodes on Qwen model family. No-G further destabilizes latency, with TTFT spikes and TPOT variance rising by 20--30\%. In contrast, full AgentServe maintains stable TTFT and TPOT across 3B/7B models and both A5000/5090 GPUs. The same trend holds for LLaMA3-8B: on the A5000, removing either component increases TTFT by 500--1000 ms and TPOT by 5--10 ms. On the 5090, both ablations still degrade TTFT by more than 100 ms and inflate TPOT by 2--4 ms. These results demonstrate the complementary benefits of adaptive scheduling and GPU resource isolation. Since No-A disables both adaptive controls, the observed gains reflect their combined effect rather than the isolated contribution of either variable. These ablations also clarify the terms in our competitive-ratio bound: removing TPOT-driven control effectively increases reservation overshoot and lag (larger $\delta$), while removing Green Context isolation weakens the discrete allocation guarantee and increases runtime interference, thereby reducing the retained fraction of prefill service.

We further stress-test AgentServe and all baselines on RTX 5090, where the larger memory headroom permits stress settings beyond the A5000 boundary. These settings use 8192-token maximum contexts for both Qwen2.5-7B and LLaMA3-8B, and the longer cold prefills are constructed by adding more tool/schema information and more complex task instructions rather than injecting artificial noise. We vary concurrency from six to eight agents to examine whether TPOT remains stable under realistic model, prompt, and concurrency pressure.

\begin{table}[t]
\centering
\caption{Stress test on RTX 5090. Entries report TTFT$_{95}$/TPOT$_{95}$ and Thr./SLO, with latencies in ms.}
\label{tab:stress-test}
\scriptsize
\setlength{\tabcolsep}{1.4pt}
\renewcommand{\arraystretch}{0.98}
\begin{tabularx}{\columnwidth}{
>{\raggedright\arraybackslash}p{0.21\columnwidth}
>{\centering\arraybackslash}X
>{\centering\arraybackslash}X
>{\centering\arraybackslash}X
>{\centering\arraybackslash}X
}
\toprule
Setting 
& AgentServe 
& SGLang 
& vLLM 
& llama.cpp \\
\midrule
Qwen-7B, 6A-8K
& \begin{tabular}{@{}c@{}}1040/27\\88/0.84\end{tabular}
& \begin{tabular}{@{}c@{}}1420/39\\61/0.59\end{tabular}
& \begin{tabular}{@{}c@{}}2180/58\\48/0.36\end{tabular}
& \begin{tabular}{@{}c@{}}3420/88\\27/0.20\end{tabular} \\
\midrule
Qwen-7B, 8A-8K
& \begin{tabular}{@{}c@{}}1450/36\\82/0.76\end{tabular}
& \begin{tabular}{@{}c@{}}2050/54\\52/0.46\end{tabular}
& \begin{tabular}{@{}c@{}}3100/82\\41/0.27\end{tabular}
& \begin{tabular}{@{}c@{}}4920/126\\22/0.14\end{tabular} \\
\midrule
LLaMA-8B, 6A-8K
& \begin{tabular}{@{}c@{}}1780/45\\72/0.68\end{tabular}
& \begin{tabular}{@{}c@{}}2520/68\\47/0.40\end{tabular}
& \begin{tabular}{@{}c@{}}3720/101\\34/0.24\end{tabular}
& \begin{tabular}{@{}c@{}}5980/150\\18/0.10\end{tabular} \\
\midrule
LLaMA-8B, 8A-8K
& \begin{tabular}{@{}c@{}}2420/58\\61/0.57\end{tabular}
& \begin{tabular}{@{}c@{}}3560/86\\34/0.27\end{tabular}
& \begin{tabular}{@{}c@{}}5300/128\\24/0.15\end{tabular}
& \begin{tabular}{@{}c@{}}7600/188\\12/0.06\end{tabular} \\
\bottomrule
\end{tabularx}
\end{table}

Table~\ref{tab:stress-test} shows that all systems degrade as the workload moves to longer tool-rich prefills, more agents, and the larger LLaMA3-8B model. However, AgentServe's TPOT increases gradually rather than collapsing, and it retains the lowest p95 TTFT/TPOT, the highest throughput, and the best SLO attainment in every stress setting. The baselines suffer larger tail-latency increases and goodput loss, indicating that phase-aware isolation keeps the TPOT feedback loop stable under realistic extreme serving conditions.
\subsection{Token Distribution Analysis and System Overhead}\label{sec:token-dist}
\begin{table}[h]
\centering
\caption{Token distribution across workloads and models. 
For prefills, the numbers indicate input tokens, and for decodes, the numbers indicate output tokens. 
The ranges represent minimum to maximum values, and the values in parentheses represent the averages.}
\label{tab:token-dist}
\resizebox{0.94\linewidth}{!}{
\begin{tabular}{l| l| ccc}
\toprule
 & \textbf{Stage} & \textbf{Qwen2.5-3B} & \textbf{Qwen2.5-7B} & \textbf{Llama3-8B} \\
\hline
\multirow{3}{*}{\rotatebox[origin=c]{90}{ReAct}} 
  & Cold Prefill   & \multicolumn{3}{c}{2.5k--3.5k tokens} \\
   \cline{2-5}
  & Resume Prefill & \multicolumn{3}{c}{30--127(56)} \\
  \cline{2-5}
  & Decode         & 27-99(37) & 21-127(45) & 32-101(38) \\
\midrule
\multirow{3}{*}{\rotatebox[origin=c]{90}{P\&E}} 
  & Cold Prefill   & \multicolumn{3}{c}{2.5k--3.5k tokens} \\
  \cline{2-5}
  & Resume Prefill & \multicolumn{3}{c}{125--421(251)} \\
  \cline{2-5}
  & Decode         & 41-125(55) & 33-141(62) & 22--116(64) \\
\bottomrule
\end{tabular}
}
\end{table}
In this section we provide token-level statistics for the two representative workloads, ReAct and Plan-and-Execute, 
evaluated on three models: Qwen2.5-3B, Qwen2.5-7B, and LLaMA3-8B. 
The purpose of this experiment is to verify empirically that agent workloads can be decomposed into three distinct execution modes, namely cold prefill, resume prefill, and decode, which are substantially different in their token scales. A second purpose is to demonstrate that these structural differences are consistent across models of different sizes.  

The results reveal several clear patterns. 
Cold prefills are primarily composed of the system prompt and tool descriptions, and therefore their length is relatively fixed. They always correspond to the longest input sequences, with lengths around 2.5k to 3.5k tokens, and they dominate the overall input regardless of workload type.
Resume prefills are shorter in length, but in the ReAct workload they appear frequently with lengths between 30 and 127 tokens (average 56), while in the Plan-and-Execute workload they occur less often but are considerably longer, ranging from 125 to 421 tokens (average 251). 
Decodes are consistently short outputs. In ReAct the decodes are extremely short, typically a few dozen tokens, whereas in Plan-and-Execute the decodes are moderately longer, between several dozen and just over one hundred tokens.  

These findings validate the workload characterization discussed in the main body of the paper. 
Cold prefills introduce risks of head-of-line blocking, frequent resume prefills interfere with the stability of decode performance, and short decodes amplify sensitivity to latency variations. 
The distributions therefore provide quantitative evidence that motivates the design of the isolation and scheduling mechanisms in AgentServe.

\textbf{System overhead.}
The TPOT p95 results in Section~IV-B directly measure serving-level token-emission stability under contention. Here, we further measure the runtime overhead introduced by TPOT monitoring and the scheduling control loop. As shown in Table~\ref{tab:system-overhead}, TPOT aggregation and scheduler decisions each take less than 1~$\mu$s per control window, and Green Context rebinding costs 6--40~$\mu$s per switch. The amortized control overhead is only 1.667~$\mu$s per window, which is several orders of magnitude smaller than the millisecond-level TPOT reported in our evaluation. Therefore, the reported TPOT stability is not obtained by ignoring feedback or control overhead. These costs are already included in the end-to-end measurements.

\begin{table}[t]
\centering
\caption{Measured runtime overhead of TPOT monitoring and scheduling control.}
\label{tab:system-overhead}
\footnotesize
\setlength{\tabcolsep}{4pt}
\renewcommand{\arraystretch}{1.08}
\begin{tabular}{p{0.56\linewidth}p{0.34\linewidth}}
\toprule
Metric & Value \\
\midrule
TPOT measurement 
& 0.017~$\mu$s/token \\
TPOT aggregation 
& $<1$~$\mu$s/control window \\
Scheduler decision 
& $<1$~$\mu$s/control window \\
Queue update / context-switch decision 
& 22.286~$\mu$s/scheduler call \\
Green Context rebinding 
& 6--40~$\mu$s/switch \\
\midrule
Amortized total control overhead 
& 1.667~$\mu$s/control window \\
\bottomrule
\end{tabular}
\end{table}
\section{Related Work}

\textbf{LLM Inference Serving.} vLLM with PagedAttention introduces paging-style memory allocation to address fragmentation and enable continuous batching \cite{kwon2023efficient}. RadixAttention in SGLang reuses cached prefixes across multi-turn and tool-invoking workloads \cite{NEURIPS2024_724be447}, while StreamingLLM proposes rolling retention for streaming contexts \cite{xiao2024efficient}. Other prefix caching methods such as CacheGen\cite{liu2024cachegen}, Marconi\cite{pan2025marconi}, and Specache\cite{jie2025specache} accelerate repeated prompts and reduce redundant computation. InfiniGen and FixGen selectively prefetch KV states and model weights to reduce CPU–GPU transfer overhead \cite{lee2024infinigen,11298528}, and vAttention introduces dynamic memory management for long-context decoding \cite{prabhu2025vattention}. Complementary quantization schemes such as GPTQ, and AWQ compress KV storage without significant quality loss\cite{frantar2023optq,lin2024awq}. Together, these methods address memory, computation, and bandwidth bottlenecks for scalable serving.

\textbf{Prefill and Decode Optimization Methods.}  
A second line of work focuses on the heterogeneity of prefill and decode. Sarathi-Serve proposes chunked prefill to mitigate head-of-line blocking \cite{298679}, while FastServe uses token-level preemption to reduce latency tails\cite{wu2023fast}. Duoserve-Moe\cite{zhang2025duoserve} employs different scheduling approaches to handle expert models during prefill and decode. DistServe disaggregates prefill and decode across GPUs \cite{298687}, and Nexus explores intra-GPU disaggregation to avoid inter-device overhead\cite{shi2025nexus}. POD-Attention introduces adaptive kernels and hybrid scheduling for mixed workloads \cite{kamath2025pod}, and FlashDecoding++ demonstrates asynchronous decoding pipelines for better per-device efficiency \cite{hong2024flashdecoding++}.  Although effective at the cluster scale, these approaches are less suitable for single-GPU agent workloads where requests are short and tightly interleaved.

\textbf{Agentic AI and System-level Approaches.}  
Recent research has examined supporting LLM agents that interleave inference with tool use. Reasoning paradigms include ReAct, Plan-and-Execute, and multi-agent collaboration \cite{yao2023react,wang-etal-2023-plan,madaan2023self}. Benchmarks such as AgentBench evaluate reasoning and retrieval-intensive tasks \cite{liu2024agentbench}. Frameworks like AutoGen, LangChain, and LangGraph enable structured multi-agent orchestration \cite{wu2024autogen,langchain2023}, while Autellix optimizes agent scheduling as general program execution \cite{luo2025autellix}. System-level support has also been studied for retrieval-augmented generation through methods like RAGCache and CacheBlend \cite{jin2024ragcache,yao2025cacheblend}. These studies highlight infrastructure adaptations for agent and RAG paradigms that complement the challenges targeted by AgentServe.
\section{Limitations}
AgentServe targets local tool-augmented SLM agents on a single consumer-grade GPU, where privacy, cost, or on-device constraints favor small and medium models over cloud-scale LLMs. It focuses on serving stability for structured tool-assisted workflows rather than claiming to support arbitrarily complex reasoning tasks.

The current design assumes sequential reasoning--action loops that alternate between model execution and tool calls, as in ReAct and Plan-and-Execute. Highly parallel or nested tool calls may introduce dependencies not explicitly modeled here. Future work could explore dependency-aware scheduling to support such workflows. For very large models, extremely long contexts, or excessive agent counts, GPU memory and single-device throughput can become the dominant bottlenecks. AgentServe protects decode responsiveness when the model and KV cache fit in memory, but it does not remove the fundamental capacity limits of model weights, KV cache, and saturated prefill throughput. Future extensions could combine memory-aware scheduling with multi-GPU execution to accommodate larger workloads.
\section{Conclusion}
This paper presented AgentServe, a single-GPU inference serving system designed for agentic workloads. By combining request isolation, TPOT-driven scheduling, and GPU-level resource partitioning, AgentServe achieves stable latency and efficient throughput under high concurrency. The design isolates cold prefills into dedicated queues, regulates resume prefills under a dynamic token budget, and enforces decode priority through fine-grained SM partitioning with CUDA Green Contexts. Our evaluation demonstrates that the system sustains scalability across different model sizes and hardware generations, delivering up to 2.8$\times$ TTFT improvement and 2.7$\times$ TPOT improvement over state-of-the-art baselines across different experimental settings.
\bibliographystyle{IEEEtran}
\bibliography{ref}

\begin{IEEEbiography}
[{\includegraphics[width=1in,height=1.25in,clip,keepaspectratio]{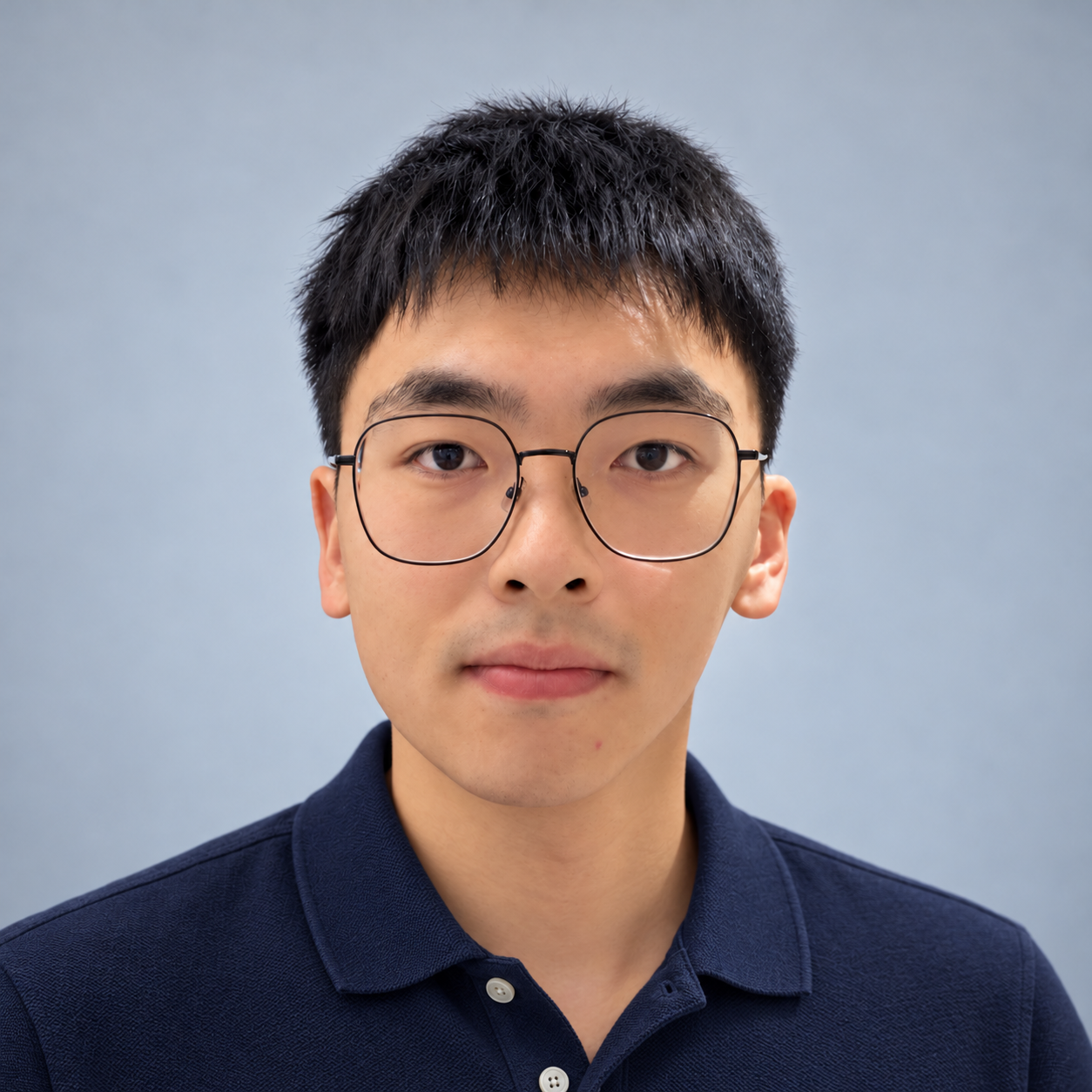}}]{Yuning Zhang received his B.Eng. degree and M. Phil degree in Electrical Engineering from the University of Sydney, Sydney, Australia, in 2021 and 2023. He is currently pursuing a PhD at the School of Computer and Electrical Engineering, the University of Sydney, with a research focus on systems for machine learning.}
\end{IEEEbiography}
\vspace{-0.8\baselineskip}
\begin{IEEEbiography}
[{\includegraphics[width=1in,height=1.25in,clip,keepaspectratio]{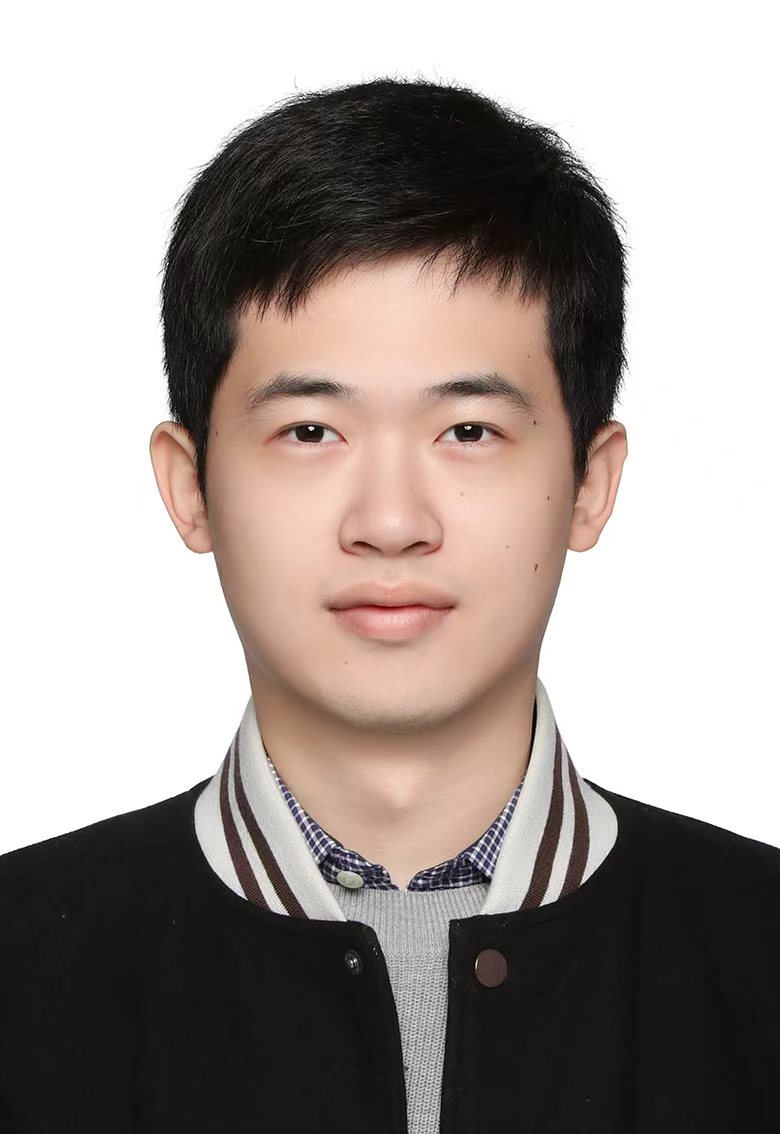}}]{Yan Yan received his Bachelor of Science degrees in Computer Science and Mathematics from Purdue University, West Lafayette, IN, USA, in 2022, where he graduated with honours and a concentration in Machine Intelligence and Security. He is currently pursuing his Master of Computer Science with a specialization in Software Engineering and Algorithm at the University of Sydney, Sydney, Australia, expected to graduate in 2026.}
\end{IEEEbiography}
\vspace{-0.8\baselineskip}
\begin{IEEEbiography}
[{\includegraphics[width=1in,height=1.25in,clip,keepaspectratio]{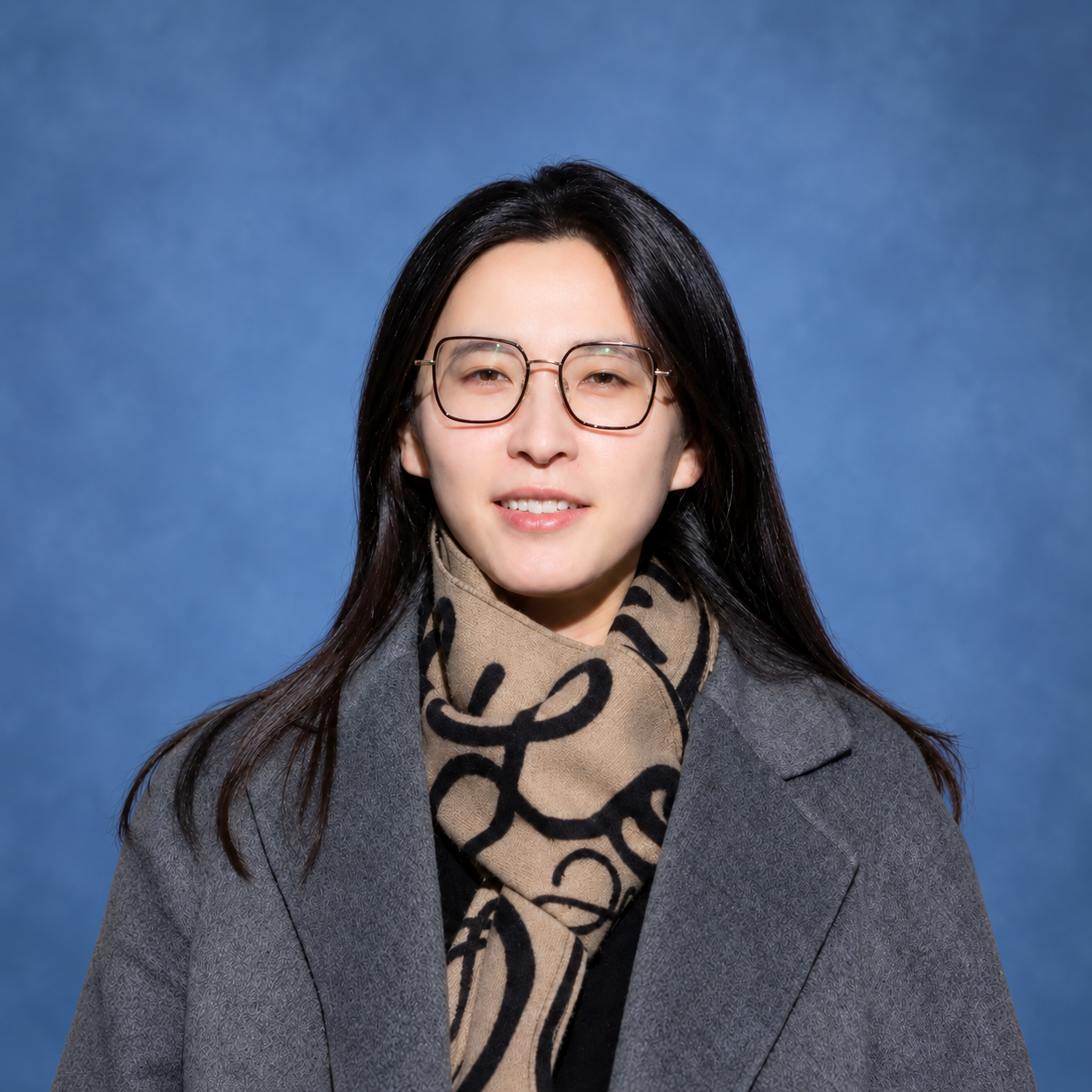}}]{Nan Yang is an Adjunct Lecturer at Western Sydney University and a Lecturer at the University of Southern Queensland. She received her B.Eng. in Software Engineering from Huaqiao University in 2015, and her M.Eng. in Electrical Engineering and PhD in Computer Science from the University of Sydney in 2018 and 2024, respectively. Her research interests include AI, distributed systems, federated learning, and cloud computing.}
\end{IEEEbiography}
\vspace{-0.8\baselineskip}
\begin{IEEEbiography}
[{\includegraphics[width=1in,height=1.25in,clip,keepaspectratio]{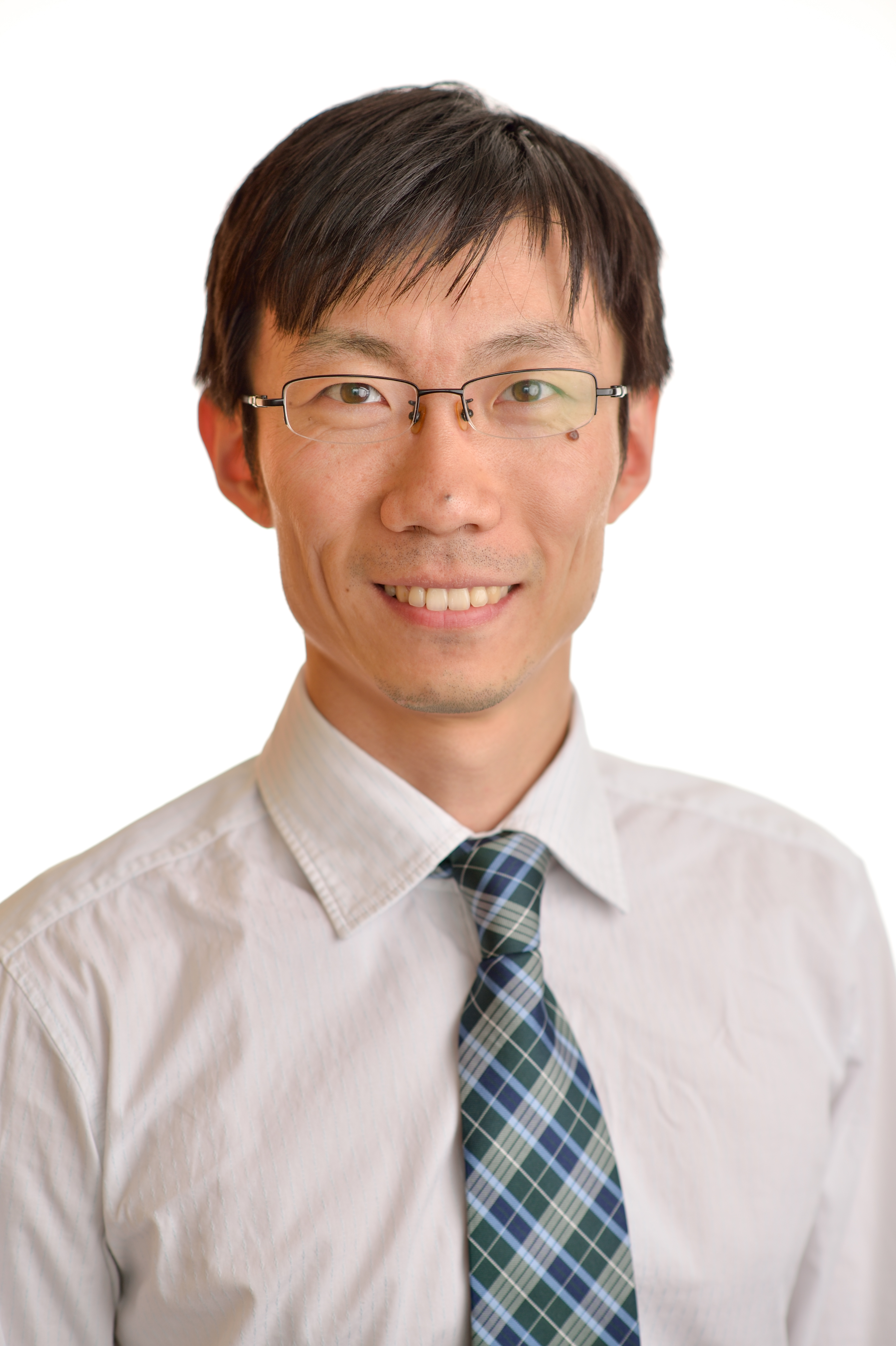}}]{Dong Yuan is an Associate Professor at the School of Electrical and Computer Engineering, the University of Sydney. He received the B. Eng. degree and M. Eng. degree from Shandong University, Jinan, China, in 2005 and 2008, the PhD degree from Swinburne University of Technology, Melbourne, Australia, in 2012, all in computer science. His research interests include digital twin, wireless sensing, artificial intelligence, cloud and edge computing, and cybersecurity.}
\end{IEEEbiography}
\end{document}